\documentclass[10pt,twocolumn,english]{article}
\usepackage{lmodern}
\usepackage[latin9]{inputenc}
\usepackage{geometry}
\usepackage{array}
\usepackage{booktabs}
\usepackage{units}
\usepackage{url}
\usepackage{amsmath}
\usepackage{amssymb}
\usepackage{wasysym}
\usepackage[authoryear]{natbib}
\usepackage{microtype}

\makeatletter

\DeclareTextSymbolDefault{\textquotedbl}{T1}
\providecommand{\tabularnewline}{\\}

\newcommand{\lyxaddress}[1]{
	\par {\raggedright #1
	\vspace{1.4em}
	\noindent\par}
}

\usepackage{longtable} 
\usepackage{widetext}
\usepackage{tikz}
\renewcommand{\pgfimage}[2][]{\includegraphics[#1]{#2}}

\usepackage{letltxmacro}
\LetLtxMacro\latexincludegraphics\includegraphics
\renewcommand{\includegraphics}[2][]{\centering\latexincludegraphics[#1]{#2}}

\bibpunct{(}{)}{;}{a}{}{,} % to follow the A&A style

\@ifundefined{showcaptionsetup}{}{%
 \PassOptionsToPackage{caption=false}{subfig}}
\usepackage{subfig}
\makeatother

\usepackage{babel}
\usepackage{listings}
\begin{document}
\global\long\def\diff{\mathop{}\!\mathrm{d}}%

\global\long\def\e{\mathrm{e}}%

\global\long\def\im{\mathrm{i}}%

\global\long\def\pvec#1{\vec{#1}\mkern2mu  \vphantom{#1}}%

\binoppenalty=10000
\relpenalty=10000
%\titlerunning{The Multivariate Extension of the Lomb-Scargle Method}
\title{The Multivariate Extension of the Lomb-Scargle Method with its technical
and astrophysical application}
\author{Martin Seilmayer \and  Ferran Garcia Gonzalez \and  Thomas Wondrak}
\maketitle

\lyxaddress{Helmholtz-Zentrum Dresden - Rossendorf e.V., Institut für Fluiddynamik,
Bautzner Landstraße 400, 01328 Dresden\\
m.seilmayer@hzdr.de}
\begin{abstract}
The common methods of spectral analysis for multivariate ($n$-dimensional)
time series, like discrete Fourier transform (FT) or Wavelet transform,
are based on Fourier series to decompose discrete data into a set
of trigonometric model components. The finite range of discrete data
causes several limitations that originate from the orthogonality mismatch
of the trigonometric basis functions on a finite interval. In the
general situation of non-equidistant or fragmented sampling, FT based
methods will cause significant errors in the parameter estimation.
Therefore, the classical Lomb\textendash Scargle method (LSM), which
is not based on Fourier series decomposition, was developed as a statistical
tool for one dimensional data. The present work generalizes LSM for
$n$-dimensional data sets by a redefinition of the shifting parameter
$\tau$, to maintain orthogonality of the trigonometric basis. An
analytical derivation shows that $n$-dimensional LSM extents the
traditional 1D case preserving all the statistical benefits, such
as consistency or the improved noise reduction. Here, we derive the
parameter confidence intervals for LSM and compare it with FT. Astrophysical
and experimental applications as well as ideal test data will illustrate
and support the proposed method.
\end{abstract}
Methods: data analysis, Techniques: spectroscopic, Sun: sunspots

\section{Introduction}

In many signal processing applications the power spectrum, or the
amplitude and phase spectrum, of a physical process is of interest.
Usually, the recorded signal is sampled in a finite equidistant time
interval and stored as discrete data values. Typical applications
of spectral analysis are the determination of characteristic frequencies,
the appropriate choice of features in terms of frequency selection
or suppression, as well as the precise measurement of amplitude and
phase of a specific frequency. In most cases, such techniques rely
on the discrete Fourier transform (DFT) or its fast version, the fast
Fourier transformation (FFT). Both utilize the Fourier series to decompose
the data into a set of coefficients from which the power spectrum
can be deduced \citep{cohen1995timefrequency,Oppenheim1999}. These
transformations are invertible, fast and easy to implement, but require
equidistant sampling of the signal which is a hard restriction for
scientific data, e.\,g. in astrophysics. From the well known one-dimensional
case, DFT can be easily extended to higher dimensions. Here the image
reconstruction for magneto-resonance tomography \citep{Haacke1999},
image filtering \citep{Gonzalez2008} or higher order spectral filters
in space and time \citep{James2011} represent just a few examples.
Non-uniform sampling is utilized in the ultra fast nuclear magnetic
resonance spectroscopy to gain a significant speed up for chemical
analysis, see \citet{frydman2003principles,giraudeau2014ultrafast}.
The cost of introducing a significant amount of missing values to
a DFT procedure is a reduced sensitivity and loss of signal amplitude.
The latter is a natural phenomenon of DFT or related procedures which
deal with incomplete data sets and gaps, see \citet{munteanu2016effectof}.
However, the established methods for general sampling schemes like
the ``non-uniform Fourier transform'', e.\,g. \citet{fessler2002iterative,fessler2003nonuniform,liu1998anaccurate},
still rely on classical orthogonal mode decomposition (application
of FFT, DFT, Wavelet transform, etc.) with its systematical errors,
as will be shown in Section~\ref{subsec:Trigonometric-orthogonal-basis}.

The application of techniques based on Fourier series to signals with
non-equidistant sampling is difficult, since it usually requires zero
padding (replacing missing values by zeros) on a grid and in the general
case resampling or interpolation of the data to a regular grid \citep{PressRybicki1989,greengard2004accelerating}.
The effect of a gap filling method is investigated extensively by
\citet{munteanu2016effectof}. Here, the conclusion is that without
a corrective measure, errors such as amplitude minimization, which
depends on the total gap density, cannot be avoided when FFT or DFT
is applied. On the other hand, interpolation could be utilized to
shift non-uniformly sampled data to a regular grid, so that the new
interpolated data points are composed of the signal information and
a projection of the accompanying noise. In the general case such distortions
or noise are not band limited leading to biased estimates which consist
of the local and aliased errors (noise, outliers, missing values,
etc.) of the surrounding data.

Astrophysical data is often affected by gaps, missing values and unevenly
sampling. For example, when ground based radio telescopes are exploring
areas in space there might exist time intervals in which the antenna
is not pointing towards the object of interest due to the rotation
of the Earth. The result is an incomplete data set. Non-uniform (random)
sampling emerges for example if the irregular appearance of objects
like sunspots is measured as a binary quality depending on time and
location. Section~\ref{subsec:Analyzing-2D-sun} provides a working
example which analyzes frequencies and periods in latitude and time
from the two dimensional data set of sunspot observations. An other
technical example for random sampling (with highly variable sampling
frequency) is asynchronous data acquisition in large sensor networks,
which are used in Smart Home, Industry 4.0 and automated driving,
see \citet{geneva2018asynchronous,cadena2016pastpresent,sudars2010dataacquisition}.
Here the time series provide missing values and data gaps originating
from time periods, in which strong noise prevents the measurement
or simply the source of the signal to be measured is not in the range
of the sensor.

The Lomb-Scargle method (LSM) was developed especially for ground
based non-uniformly sampled one dimensional data, from which the amplitude
spectrum is calculated \citep{lomb1976leastsquares,scargle1982studies}.
The main advantage of this method is to directly estimate the spectrum
without iterative optimization of trigonometric models as discussed
in Section~\ref{subsec:leastSqareFit}. A fast version for one dimensional
signals is presented by \citet{townsend2010fastcalculation} and \citet{leroy2012fastcalculation}.

An extension of LSM to two- or three-dimensional time series have
not been presented yet. Especially for large multivariate data sets
such a direct approach would be comparably faster than present iterative
procedures as proposed in \citet[Chap. 9]{babu2010spectral}. The
advantages of multivariate LSM will be demonstrated in Section~\ref{sec:Application}
by means of a 3D Ultrasound flow profile measurement and the 2D analysis
of sunspot time series data.

In case of the analysis of flow profile measurements which are obtained
in an experiment investigating the magnetorotational instability in
liquid metals \citep{seilmayer2014experimental}, such method would
be highly desirable. Due to the complex experimental setup and the
weak signal to noise ratio, the flow profile measurements contain
several time intervals, in which distortions are dominant, see \citet{seilmayer2016}.
These time intervals had to be rejected leading to a time series with
invalid (missing) data points, from which the multivariate version
of LSM is able to determine the parameters of the characteristic traveling
wave.

In order to demonstrate the method and to motivate the basic idea
of the multivariate version of LSM it is compared with the traditional
orthogonal mode decomposition (OMD) with trigonometric basis functions,
which is the essence of the classical Fourier transform, in terms
of necessary conditions and error (noise) behavior. It will be shown
that LSM fits better to the conditions of an arbitrary finite length
of the sampling series in comparison to DFT, because LSM reduces the
error in model parameter estimation. In contrast to the traditional
approach, which was derived from a statistical point of view (see
the appendix in J.\,D.~Scargle \citet{scargle1982studies}), the
presented method is deduced from a technical point of view and focuses
on its application. This leads to a slight change in the scaling of
model parameters, which will be discussed in Section~\ref{subsec:Lombs-Idea}.
However, the introduced procedure includes all benefits, like arbitrary
sampling, fragmented data and a good noise rejection.

The starting point of this paper is the analysis of a continuous 1D
signal $s:\mathbb{R\rightarrow R}$ which is composed of an arbitrary
and finite set of individual frequency components $\omega_{i},0\leq i\leq M$.
Without loss of generality, band limitation is assumed stating that
there exists an upper maximum frequency $\omega_{\max}$ with $\omega_{i}<\omega_{\max},0\leq i\leq M$,
which mimics an intrinsic low pass filter characteristic of the measurement
device. Since the measurement time is finite, the value of $s$ is
only known in the time interval $\left[0,T\right]$ with $T\in\mathbb{R}$
and $T>0$. The $m$-dimensional extension of this signal is $S:\mathbb{R^{\mathrm{\mathit{m}}}\rightarrow R}$.
If such a continuous signal is sampled, the pair $(\hat{s}_{i},t_{i})\in\mathbb{R}^{2},0\leq i\leq N-1,$
represents the measured 1D value and the corresponding instant in
time whereas the pair $(\hat{S}_{i},\vec{t}_{i})\in\mathbb{R}^{m+1},0\leq i\leq N-1,$
represents a measured value and the $m$-dimensional location (space/time)
for the $i$-th sample ($N$ is the number of samples). All the methods
presented in this paper are implemented in a package written in \textsf{R}
and published on CRAN \citep{seilmayer2019commonmethods}.

\section{Mathematical model and comparison of OMD and LSM}

In order to delineate the differences between trigonometric OMD and
LSM, we start with the basic model of a periodic signal as a sum of
signals of different frequencies $\omega_{k},0\leq k\leq M,$ with
the corresponding amplitude $A_{k}\in\mathbb{R}$ and phase shift
$\varphi_{k}\in\mathbb{R}$. The corresponding trigonometric model
function 
\begin{align}
y(t) & =\sum_{k=0}^{M}A_{k}\cos(\omega_{k}t+\varphi_{k})\label{eq:QDT_model-by-phase}\\
 & =\sum_{k=0}^{M}\Bigl(a_{k}\cos(\omega_{k}t)+b_{k}\sin(\omega_{k}t)\Bigl)\label{eq:QDT_model-1}
\end{align}
describes an infinite, stationary and steady process $y:\mathbb{R}\rightarrow\mathbb{R}$
with the coefficients $a_{k},b_{k}\in\mathbb{R}$ for the defined
frequency $\omega_{k}$ and the identities $A_{k}=\sqrt{a_{k}^{2}+b_{k}^{2}}$
as well as $\varphi_{k}=\tan^{-1}(b_{k}/a_{k})$. Furthermore, the
trigonometric model above consists of an arbitrary number $M\in\mathbb{N}_{0}$
of frequency components. If the given signal $s(t)$ is described
by the defined model from Eq.~(\ref{eq:QDT_model-1}) the model misfit
$\epsilon$ is given by
\begin{equation}
\epsilon(t)=s(t)-y(t).\label{eq:model-misfit}
\end{equation}
Thus, the signal can now be described by inserting Eq.~(\ref{eq:QDT_model-1})
into Eq.~(\ref{eq:model-misfit}) and defining the misfit $\epsilon_{k}$
for each discrete frequency $k$ with $\epsilon(t)=\sum_{k}\epsilon_{k}(t)$
as follows:
\begin{align}
s(t) & =\sum_{k=0}^{M}\Bigl(a_{k}\cos(\omega_{k}t)+b_{k}\sin(\omega_{k}t)+\epsilon_{k}(t)\Bigr).\label{eq:ModelWithError}
\end{align}
The defined misfit originates form measurement uncertainties or parametric
errors from $a_{k}$, $b_{k}$. In the general case $\epsilon(t)$
can be any function or distribution. The challenge is to precisely
determine the model parameters $a_{k}$ and $b_{k}$ for a given signal
$s(t)$ achieving minimal $\epsilon(t)$. This can be accomplished
by one of the following three methods: (i) Least-Square fit; (ii)
orthogonal mode decomposition; (iii) Lomb-Scargle method (LSM).

\subsection{Approach i \textendash{} Least square fit}

\label{subsec:leastSqareFit}The optimal fit is reached by least square
fitting, resulting in a minimum $\epsilon$, which was shown by \citet{barning1963thenumerical,mathias2004algorithms}.
Since such procedures are iterative, the convergence of the algorithm
might need a large number of function evaluations of Eq.~(\ref{eq:ModelWithError}).
Therefore, a direct version like LSM is preferable. It can be shown
that LSM becomes equivalent to a least square fit of a sinusoidal
model \citep{lomb1976leastsquares,barning1963thenumerical}.

\subsection{Approach ii \textendash{} Trigonometric OMD}

\label{subsec:Trigonometric-orthogonal-basis}Generally, two functions
$f,g:\mathbb{R}\rightarrow\mathbb{R}$ are said to be orthogonal on
the interval $\left[a,b\right]\subset\mathbb{R},$ if the following
condition holds, refer to \citet{weisstein2019orthogonal}:
\begin{equation}
\int_{a}^{b}f(x)g(x)\diff x=0.\label{eq:FunctionOrthogonality}
\end{equation}
By selecting $f(x)=\sin(x)$ and $g(x)=\cos(x)$, the integration
leads, by exploiting the identity $\cos(x)\sin(x)=\frac{1}{2}\sin(2x)$,
to
\begin{equation}
\frac{1}{4}\Bigl(\cos(2a)-\cos(2b)\Bigl)=0,\label{eq:FunctionOrthogonalityTrig}
\end{equation}
which is only zero, if $\cos(2a)=\cos(2b).$ This is true for any
$a\in\mathbb{R}$, if the length of the interval $[a,b]$ is a multiple
of the period $2\pi$ so that $2b=2a+2\pi k$ with $k\in\mathbb{N}^{+}$.
It is interesting to note that this interval can be shorted to one
half of the period, if $\cos(2a)$ is zero. By exploiting this feature
of the trigonometric functions, the individual model coefficients
are calculated by multiplying the sine or the cosine to the measured
data $s(t)$ and integrating over all times as shown by \citet[Chap. 15]{cohen1995timefrequency}:
\begin{align}
a_{k} & =\frac{2}{T}\int_{-\infty}^{\infty}s(t)\cos(\omega_{k}t)\diff t\label{eq:QDT_ak-1}\\
b_{k} & =\frac{2}{T}\int_{-\infty}^{\infty}s(t)\sin(\omega_{k}t)\diff t.\label{eq:QDT_bk-1}
\end{align}
For a measured signal, the integration can only be performed over
the interval $\left[0,T\right]$. It is obvious that an error is introduced,
if $T\neq2\pi n$. In order to investigate the properties of the finite
integration, we will concentrate in the following on the cosine term
(Eq.~(\ref{eq:QDT_ak-1})), since the analysis of the sine term (Eq.~(\ref{eq:QDT_bk-1}))
is similar. By setting the integral boundaries to the finite time
interval the integral in Eq.~(\ref{eq:QDT_ak-1}) can be written
as

\begin{align}
\int_{0}^{T}s(t)\cos(\omega_{k}t)dt= & \int_{0}^{T}\Bigl(a_{k}\cos{{}^2}(\omega_{k}t)\label{eq:QDT-pre}\\
 & +b_{k}\sin(\omega_{k}t)\cos(\omega_{k}t)\nonumber \\
 & +\epsilon_{k}(t)\cos(\omega_{k}t)\Bigl)\,\diff t.\nonumber 
\end{align}

With the two identities $\cos^{2}(x)=\frac{1}{2}\left(1+\cos(2x)\right)$
and $\sin(x)\cos(x)=\frac{1}{2}\sin(2x)$ the integral (\ref{eq:QDT-pre})
is then expanded to

\begin{multline}
\int_{0}^{T}s(t)\cos(\omega_{k}t)\diff t\\
=\frac{a_{k}}{2}\int_{0}^{T}\Bigl(1+\cos(2\omega_{k}t)+2\frac{b_{k}}{a_{k}}\sin(2\omega_{k}t)\cos(\omega_{k}t)\Bigl)\diff t\\
+\int_{0}^{T}\epsilon_{k}(t)\cos(\omega_{k}t)\diff t\label{eq:OBD_ContinuousStart}
\end{multline}
and further reduced to

\begin{multline}
\frac{2}{T}\int_{0}^{T}s(t)\cos(\omega_{k}t)\diff t\\
=a_{k}\Biggl(1+\underset{\text{truncation error }\epsilon_{\mathrm{Tk}}}{\underbrace{\frac{1}{T}\int_{0}^{T}\Bigl(\cos(2\omega_{k}t)+\frac{b_{k}}{a_{k}}\sin(2\omega_{k}t)\Bigl)\diff t}\Biggl)}\\
+\underset{\text{random error }\epsilon_{\mathrm{FSk}}}{\underbrace{\frac{2}{T}\int_{0}^{T}\epsilon_{k}(t)\cos(\omega_{k}t)\diff t}}.\label{eq:OBD_Continuous}
\end{multline}

The equation above indicates that the coefficient $a_{k}$ is effected
by two errors $\epsilon_{\mathrm{T}k}$ (truncation) and $\epsilon_{\mathrm{FS}k}$
(random) which might be nonzero for an arbitrary $T$. Thus, if both
errors are neglected, the integral on the left hand side turns into
an approximation of $a_{k}$. Taking the consideration about the orthogonality
described in Eq.~(\ref{eq:FunctionOrthogonality}) into account,
$\epsilon_{\mathrm{T}k}$ becomes only zero, if the integration time
$T$ is an integer multiple of $\pi/\omega_{k}$. This means for the
Fourier series that a given integration time $T$ (observation window)
defines the lowest allowed frequency $\omega_{0}$. Furthermore, only
multiples of this fundamental frequency $\omega_{k}=k\omega_{0}$
with $k\in\mathbb{N}$ are allowed in the Fourier series because $\epsilon_{\mathrm{T}k}=0$
in this case.

The technical realization of OMD is called quadrature demodulation,
which is the discrete version of equation~(\ref{eq:OBD_Continuous})
by replacing the integral over $s(t)$ into a sum over the discrete
sampled values $\hat{s}_{n},0\leq n\leq N-1$ and setting the truncation
error to zero. For a time discrete signal with $N$ samples and equidistant
sampling with a constant sampling period $T_{s}$, and $t_{n}=nT_{s}$,
the parameter $a_{k}$ is approximated for a certain frequency $\omega_{k}$
by 
\begin{equation}
a_{k}\approx\frac{2}{N}\sum_{n=0}^{N-1}\hat{s}_{n}\cos(\omega_{k}t_{n}).\label{eq:QuadratureDemodulation}
\end{equation}

According to the previous considerations, the truncation error $\epsilon_{\mathrm{T}k}$
is larger than zero, if the measurement range $T=NT_{\mathrm{s}}$
is not exactly a multiple of $\pi/\omega_{k}$ as shown in Figure~(\ref{fig:Sinus-integration}).
\begin{figure*}
\begin{centering}
\begin{center}
\input{Pictures/sin_integration.tex}
\par\end{center}
\par\end{centering}
\caption{\label{fig:Sinus-integration}Example of the truncation error of a
signal with the time period of $2\pi$, which is generated by the
last two samples and the value is indicated by the gray shaded area.}
\end{figure*}
In this example a signal with a time period of $t=2\pi$ is sampled
with $N=23$, $T_{s}=2\pi/20$. The total integration time is $T=2\pi+1/5\pi$,
being slightly longer than the period of the signal, which is indicated
by the two additional sampling points after $t=2\pi$. The truncation
error equals the gray shaded area.

Generally speaking, the sampling error in the discrete version can
be estimated by

\begin{equation}
\frac{2}{N}\sum_{n=0}^{N-1}\hat{s}_{n}\cos(\omega_{k}t_{n})\approx a_{k}\Bigl(1+\underset{\epsilon_{\mathrm{T}k}}{\underbrace{\frac{\Delta\varphi}{T}}}\Bigl)\pm\underset{\epsilon_{\mathrm{FS}k}}{\underbrace{\Phi_{1-\alpha}\frac{2\sigma}{\sqrt{N}}}},\label{eq:QDT-discrete-error}
\end{equation}
where $\Delta\varphi=\min(T-\pi i/\omega_{k})$ with the according
$i\in\mathbb{N}$ and the random error is modeled by the $\alpha$-quantile
of the sampling error distribution $\Phi_{1-\alpha}$ related to the
underlying process with its standard deviation $\sigma$. This estimates
the confidence interval of $a_{k}$ and $b_{k}$ respectively.

The above considerations lead to following four statements, which
are derived in detail in the appendix: (i) the maximum absolute value
of the truncation error is bounded from above by $\left|\epsilon_{\mathrm{Tk}}\right|\leq0.2$
which is reached, if $T\apprge\pi/\omega_{k}$, which is consistent
with the experimental findings from \citet{thompson1980leakage};
(ii) $\epsilon_{\mathrm{T}k}$ is independent from total number of
samples $N$ for a constant time $T$, which makes the OMD a non-consistent
estimator for model parameters $a_{k}$ and $b_{k}$; (iii) by increasing
the sample rate the measurement error decreases by $\mathcal{O}(N^{-0.5})$,
but scales with twice the standard deviation of noise distribution;
(iv) DFT can be derived from equation (\ref{eq:QDT-discrete-error})
by restricting to equidistant sampling with constant $T_{\mathrm{s}}$.

Concluding, the truncation error, which is an intrinsic feature of
OMD, produces a systematic deviation from the true value depending
on the difference between the sampling time interval and the corresponding
period of the frequency of interest. In contrast to that, the random
error diminishes with increasing sampling rate $T_{\mathrm{s}}^{-1}=N/T$.
A detailed discussion on this topic can be found in \citet{jerri1977theshannon,shannon1949communication,thompson1980leakage}.
In case of multivariate and randomly sampled data the recent work
of \citet{al-ani2012evaluation} suggests two calculation schemes
estimating the Fourier transform in the general case. The continuous
time Fourier transform estimation (similar to Eq.~(\ref{eq:QuadratureDemodulation}))
takes samples with arbitrary spacing as the general approach, in contrast
to the secondly proposed discrete time Fourier transform ($t_{n}=nT_{\mathrm{s}}$
and $\omega_{k}=k\omega_{0}$) estimation scheme. Here the data is
projected onto a regular grid, which now may provide locations of
missing data. Both schemes consider the sampling pattern $t_{n}$
and its power spectrum distribution function $p(t_{n})$. However,
even for such sophisticated methods, the main conceptual drawbacks
(see points (i) and (ii)) remain, which motivates the subsequent Lomb-Scargle
method as non OMD method and its extension to multivariate data.

\subsection{Approach iii \textendash{} Lomb Scargle method}

\label{subsec:Lomb-Scargle-Estimator}As described in the previous
section, the disadvantage of OMD is that cosine and sine are not orthogonal
on arbitrary intervals. By introducing an additional parameter $\tau\in\mathbb{R}$
into Equation (\ref{eq:FunctionOrthogonality}) it will be shown that

\begin{equation}
\int_{a}^{b}\cos(x-\tau)\sin(x-\tau)\diff x=0\label{eq:LMB-modell1Freq}
\end{equation}
holds for arbitrary intervals $\left[a,b\right]$. By utilizing the
trigonometric identities, $\cos^{2}(\phi)-\sin^{2}(\phi)=\cos(2\phi)$
and $\cos(\phi)\sin(\phi)=\frac{1}{2}\sin(2\phi)$, to remove the
differences in the arguments the integral can be transformed to 
\[
\int_{a}^{b}\left(\frac{1}{2}\sin\left(2x\right)\cos\left(2\tau\right)-\frac{1}{2}\sin\left(2\tau\right)\cos\left(2x\right)\right)\diff x=0.
\]
If this integral has to be zero, the following condition has to hold

\[
\cos(2\tau)\int_{a}^{b}\sin(2x)\diff x=\sin(2\tau)\int_{a}^{b}\cos(2x)\diff x
\]
which can be transformed to

\begin{equation}
\frac{\int_{a}^{b}\sin\left(2x\right)\diff x}{\int_{a}^{b}\cos\left(2x\right)\diff x}=\tan\left(2\tau\right).\label{eq:LS_Ortho_Tau4-1}
\end{equation}
Therefore, equation (\ref{eq:LS_Ortho_Tau4-1}) calculates the parameter
$\tau$ in such a way that expression (\ref{eq:LMB-modell1Freq})
is zero again. In case of equidistant sampling, the value of $\tau$
can be directly calculated from the integration boundaries in the
following way:
\[
\tau=\frac{b-a}{2}.
\]

Based on this general consideration the time shifting parameter $\tau_{k}\in\mathbb{R},0\leq k\leq M,$
was introduced by Lomb and Scargle for frequency $\omega_{k}$ into
the model given in Equation (\ref{eq:QDT_model-1}) 
\begin{equation}
s(t)=\sum_{k=0}^{M}\left(a_{k}\cos\bigl(\omega_{k}(t-\tau_{k})\bigr)+b_{k}\sin\bigl(\omega_{k}(t-\tau_{k})\bigl)+\epsilon_{k}(t)\right)\label{eq:LS-Model}
\end{equation}
in order to remove the truncation error. The parameter $\tau_{k}$
can be calculated by 
\begin{equation}
\tan(2\omega_{k}\tau_{k})=\frac{\int_{a}^{b}\sin(2\omega_{k}t)\diff t}{\int_{a}^{b}\cos(2\omega_{k}t)\diff t}\label{eq:LS-ModelTauCont}
\end{equation}
similar to equation (\ref{eq:LS_Ortho_Tau4-1}). For the time discrete
version the integrals transform into a sum resulting in
\begin{equation}
\tan(2\omega_{k}\tau_{k})=\frac{\sum_{n=0}^{N-1}\sin(2\omega_{k}t_{n})}{\sum_{n=0}^{N-1}\cos(2\omega_{k}t_{n})}.\label{eq:LS_tau-1D}
\end{equation}
The parameters $a_{k}$ and $b_{k}$ can be determined beginning with
Equation (\ref{eq:OBD_ContinuousStart}) but factorize it by $\cos^{2}x$,
instead of expanding $\cos^{2}x$ to $\frac{1}{2}\left(1+\cos(2x)\right)$
as done in Equation (\ref{eq:OBD_Continuous}). Since the method is
applied to sampled data, the method will be delineated for the discrete
set $\hat{s}_{i}$ in the following. The data is multiplied by $\cos(\omega_{k}(t_{n}-\tau_{k}))$
resulting in the next equation for a single frequency $\omega_{k}$
with $\phi_{k}=\omega_{k}(t_{n}-\tau_{k})$ as a substitution: 
\begin{multline}
\sum_{n=0}^{N-1}\hat{s}_{n}\cos(\phi_{k})\\
=a_{k}\sum_{n=0}^{N-1}\cos^{2}(\phi_{k})+b_{k}\underbrace{\sum_{n=0}^{N-1}\sin(\phi_{k})\cos(\phi_{k})}_{\text{truncation error }\epsilon_{\mathrm{Tk}}}\\
+\underbrace{\sum_{n=0}^{N-1}\epsilon_{k}(t_{n})\cos(\phi_{k})}_{\text{\text{modulated error }\ensuremath{\epsilon_{\mathrm{LS}k}}}}.\label{eq:LSM1DStart}
\end{multline}
The sum over the term $\cos(x)\sin(x)$ vanishes because of the proper
selection of $\tau_{k}$ according to Equation (\ref{eq:LS_tau-1D}).
The sum over $\epsilon_{k}(t_{n})\cos(\phi_{k})$ describes the modulated
noise distribution function. The sum over $\cos^{2}(\phi_{k})$ is
the dominating term to calculate the value of parameter $a_{k}$ which
can be directly obtained by dividing by $\sum_{n=0}^{N-1}\cos^{2}(\phi_{k})$,
which gives
\begin{equation}
\frac{\sum_{n=0}^{N-1}\hat{s}_{n}\cos(\phi_{k})}{\sum_{n=0}^{N-1}\cos^{2}(\phi_{k})}=a_{k}+\underset{\epsilon_{\mathrm{LSk}}}{\underbrace{\frac{\sum_{n=0}^{N-1}\epsilon_{k}(t_{n})\cos(\phi_{k})}{\sum_{n=0}^{N-1}\cos^{2}(\phi_{k})}}}.\label{eq:LS_Basic_ak}
\end{equation}
The residual error term on the right hand side consists of a random
distributed part divided by a sum over the square of the cosine. In
contrast to OMD, the estimation error of the parameter $a_{k}$ only
depends on noise and is independent from the realization of sampling.

The next step is to determine $\epsilon_{\mathrm{LSk}}$ in terms
of a confidence interval with respect to $\alpha$ as it was done
for OMD in Equation (\ref{eq:QDT-discrete-error}). Given a normal
distributed error function $\epsilon_{k}\leftrightarrow\mathcal{N}(0,\sigma)$,
with expectation value of zero and standard deviation $\sigma$ independent
from time, the error can be factorized (\citet[Theorem 4A, p. 90]{parzen1962stochastic}),
leading to 
\begin{align}
\epsilon_{\mathrm{LS}} & =\Phi_{1-\alpha}\frac{\sigma}{\sqrt{N}}\underset{e_{\max}}{\underbrace{\frac{\sum_{n=0}^{N-1}\cos(\phi)}{\sum_{n=0}^{N-1}\cos^{2}(\phi)}}}\label{eq:LS_error}
\end{align}
estimating the most probable limits of $\epsilon_{\mathrm{LS}}$.
Here $\Phi_{1-\alpha}$ denotes the $\alpha$-quantil of the given
distribution function. Since the estimation error neither depend on
the frequency nor the shifting parameter, we estimate

\begin{equation}
e_{\max}=\max_{\beta}\left[\frac{\int_{0}^{\beta}\cos(\phi)\diff\phi}{\int_{0}^{\beta}\cos^{2}(\phi)\diff\phi}\right].\label{eq:LS_error_factor}
\end{equation}
Without loss of generality we set $\omega_{k}=1$ and $\tau_{k}=0$.
In order to maximize $e_{\mathrm{max}}$ for $\beta\in[0,2\pi]$,
Equation (\ref{eq:LS_error_factor}) must be integrated which results
$4\sin(\beta)/(2\beta+\sin(2\beta))$ so that the maximum error is
$e_{\mathrm{max}}=\frac{4}{\pi}$ for $\beta=\pi/2$. The final parameter
estimation gives

\begin{equation}
a_{k}=\frac{\sum_{n=0}^{N-1}\hat{s}_{n}\cos(\omega_{k}(t_{n}-\tau_{k}))}{\sum_{n=0}^{N-1}\cos^{2}(\omega_{k}(t_{n}-\tau_{k}))}\pm\underset{\Delta a_{k}=\Delta b_{k}}{\underbrace{\frac{4}{\pi}\Phi_{1-\alpha}\frac{\sigma}{\sqrt{N}}}},\label{eq:LS_final}
\end{equation}
where the error converges to zero by increasing the number of samples
$N$ for a fixed time interval. This property qualifies LSM to be
a consistent estimator in amplitude and phase \citep{mathias2004algorithms},
because by increasing the number of samples the estimation gets more
precise.

In comparison to the classical definition of \citet{lomb1976leastsquares}
the coefficients $a_{k}$ and $b_{k}$ differ by a factor of $\sqrt{N/2}$
scaling the result. With a closer look to the original equations
\begin{align*}
a_{k,\mathrm{orig}} & =\frac{\sum_{n}\hat{s}_{n}\cos(\omega_{k}(t_{n}-\tau_{k}))}{\sqrt{\frac{N}{2}}\sqrt{\sum_{n}\cos^{2}(\omega_{k}(t_{n}-\tau_{k}))}}
\end{align*}
this factor becomes evident, e.\,g. \citet{hocke1998phaseestimation}.
Since $\lim_{n\rightarrow\infty}\sum_{n}\cos^{2}(\omega_{k}(t_{n}-\tau_{k}))\approx N/2$
can be assumed, $a_{k}\approx a_{k,\mathrm{orig}}$ is valid, if $N$
fits exactly to a multiple of $\omega_{k}$. Because the presented
approach is related to the rather technical amplitude demodulation
procedure, we assume equation (\ref{eq:LS_final}) to be more accurate.

The confidence interval $\Delta a_{k}$ for parameter $a_{k}$ can
be approximated by

\[
\Delta a_{k}=\frac{4}{\pi}\Phi_{1-\alpha}\frac{\sigma}{\sqrt{N}},
\]
which is smaller than the value for OMD given in Equation~(\ref{eq:QDT-discrete-error}).
The confidence interval for the amplitude $\Delta A_{k}$ can be deduced
by propagating the error 
\begin{align}
\Delta A_{k} & =\frac{\partial A_{k}}{\partial a}\Delta a_{k}+\frac{\partial A_{k}}{\partial b}\Delta b_{k}\nonumber \\
 & =\frac{4}{\pi}\Phi_{1-\alpha}\sqrt{\frac{2}{N}}\sigma.\label{eq:Delta_A}
\end{align}
In a similar way, the confidence interval for the phase $\varphi_{k}$
can be defined by 
\begin{align}
\varphi_{k} & =\tan^{-1}\left(\frac{b_{k}}{a_{k}}\right)\pm\Delta\varphi\nonumber \\
 & =\tan^{-1}\left(\frac{b_{k}}{a_{k}}\right)\pm\frac{4}{\pi}\Phi_{1-\alpha}\sqrt{\frac{2}{N}}\frac{\sigma}{A_{k}}.
\end{align}
It is interesting to note that the confidence interval for the phase
$\varphi_{k}$ decreases when the amplitude is increasing.

In order to determine the spectrum with LSM, we assume the recorded
signal is described by many frequencies. The most significant frequency
is then represented by a peak in the frequency spectrum of a certain
width and height. The width is determined by the frequency resolution
$\Delta f$ which equals $1/T$. From this point of view the precision
of a frequency estimation changes only with observation length $T$
and seams to be independent from the number of samples $N$ and signal
quality. The quality $\Sigma$ is measured by a signal-to-noise ratio
like expression
\begin{align}
\Sigma & =\sqrt{\frac{1}{N}\sum_{n=0}^{N-1}\frac{\left(s_{n}-y_{n}\right)^{2}}{\sigma_{n}^{2}}}\approx\frac{\sqrt{\sum_{l}A_{l}^{2}}}{\sqrt{\frac{1}{N}\sum_{n=0}^{N-1}\left(\epsilon(t_{n})\right)^{2}}}\label{eq:SNR}
\end{align}
with $A_{l}$ counting the \emph{significant} amplitudes. Here $y_{n}$
denotes the fitted model and $\sigma_{n}$ the uncertainty per sample,
which is related to $\sigma$ by $\sigma=\sqrt{\frac{1}{N}\sum_{n=0}^{N-1}\sigma_{n}^{2}}$
and $\epsilon(t_{n})$ as the noise per sample. Following \citet{vanderplas2017understanding}
we apply Bayesian statistics and assume that every peak is Gaussian
shaped, i.\,e. $\e^{P(f_{\max}\pm\Delta f)}\propto\e^{-\Delta f^{2}/\left(2\sigma_{f}^{2}\right)}$.
It follows that a significant peak $A_{\max}^{2}=A^{2}(f_{\max})$
appears at $f_{\max}$, in a way that $A_{\max}^{2}/2=A^{2}(f_{\max}\pm\Delta f)$
is valid. Here, $A^{2}$ is related to the power spectral density
$P(f)\propto a_{k}^{2}+b_{k}^{2}$. The frequency uncertainty (or
standard deviation) is then given by
\[
\sigma_{f}\approx\Delta f\sqrt{\frac{2}{N\Sigma^{2}}}
\]
so that a significant peak is located in the interval $f_{\max}\pm\sigma_{f}$.
This approach suggests that increasing the number of samples in a
fixed interval $T$ enhances the precision of $f_{\max}$ by reducing
$\sigma_{f}.$ However, if the original signal contains two frequencies
with a distance in the range of $1/T$, it cannot be excluded even
for LSM that these peaks merge together in one single peak with small
$\sigma_{f}$ according to \citet{kovacs1981frequency}.

Concluding the properties of LSM: (i) there is no truncation error
$\epsilon_{\mathrm{T}}=0$; (ii) LSM provides a better noise rejection
compared to OMD, $\epsilon_{\mathrm{LS}}<\epsilon_{\mathrm{FS}}$;
(iii) the explicit sampling pattern is not of interest to work with
LSM.

\subsection{Power Spectral Density and False Alarm Probability}

\label{subsec:Power-Spectral-Density}In this work the simplest case
of uncorrelated and mean free Gaussian noise is assumed which suits
many common technical and scientific cases. From the power spectral
density (PSD) $P_{k}=\frac{N}{4\sigma_{0}^{2}}(a_{k}^{2}+b_{k}^{2})$,
with $\sigma_{0}^{2}=\sum_{n=0}^{N-1}\left(y(t_{n})-\bar{y}\right)^{2}$
as the variance of the sample, refer to \citet{hocke1998phaseestimation}
and \citet{zechmeister2009thegeneralised}, the standardized PSD is
defined by 
\begin{align}
\mathrm{psd}(\omega_{k}) & =P_{k}p_{k}\label{eq:PSD}
\end{align}
on the interval $[0,1]$, where $p_{k}$ is the standardized Gaussian
noise. Here, $P_{k}$ is similar to a signal to noise ratio given
in Equation \ref{eq:SNR} (see \citet{scargle1982studies}). Since
LSM calculates the result of a least square fit, a value of $\mathrm{psd}(\omega_{k})=1$
indicates a ``perfect'' fit to the corresponding model function.
In the case of $\mathrm{psd}(\omega_{k})=0$ no correspondence is
visible. The discussion about the presented standardization is carried
out in detail by \citet{cumming1999thelick}. The different ways to
perform the calculation of the $\mathrm{psd}$-value are briefly summarized
in \citet{zechmeister2009thegeneralised}. Additionally, a more precise
description of noise takes some effort which should be accomplished
by analyzing the measurement data or by taking additional noise measurements.
The different procedures are briefly described by \citet{cumming1999thelick}
and \citet{horne1986aprescription}.

The standardized noise level reads $p_{k}=2/(N-1)$, so the standardized
power spectral density 
\begin{equation}
\mathrm{psd}(\omega_{k})=\frac{N}{N-1}\frac{A_{k}^{2}(\omega_{k})}{2\sigma_{0}^{2}}\label{eq:normPSD}
\end{equation}
can be calculated directly from the power spectral density or amplitude.
A more sophisticated approach relies on a Bayesian estimate of $\mathrm{psd}(\omega_{k})$
which is presented in \citet{mortier2015bglsa}. For most technical
applications equation (\ref{eq:normPSD}) should be sufficient.

As a statistical measure, the probability 
\begin{equation}
\mathcal{P}(P_{k}>P_{0})=\left(1-\mathrm{psd}(\omega_{k})\right)^{\frac{N-3}{2}}\label{eq:ProbFAP}
\end{equation}
states that there is no PSD peak $P_{k}$ larger than a reference
value $P_{0}$ of the best fit. From here the statistical significance
of a single frequency $\omega_{k}$ can be deduced as the so called
false alarm probability (FAP) with
\begin{equation}
\mathrm{FAP}=\begin{cases}
1-\left(1-\mathcal{P}(P_{k}>P_{0})\right)^{M} & ,\text{if }\mathcal{P}(P_{k}>P_{0})\approx1\\
M\mathcal{P}(P_{k}>P_{0}) & ,\text{if }\mathcal{P}(P_{k}>P_{0})\ll1
\end{cases},\label{eq:FAP}
\end{equation}
where $M$ denotes the number of independent (fundamental) frequencies
present in the signal. The discussion about this degree of freedom
is very diverse in literature and is discussed for instance by \citet{vanderplas2017understanding}.
The first approach is Shannon's sampling theorem as a pragmatic and
conservative access to this topic. It states that the number of independent
frequencies is $M\approx N/2$. At the same moment a band limited
signal is required which is sampled with twice the maximum signal
frequency $f_{\mathrm{s}}\geq2\max(f)$. It follows that signal frequencies
above $f_{\mathrm{s}}/2$ become visible as an alias in the lower
frequency domain. In this aspect randomly sampled data may behave
different. For randomly sampled data the conservative approach defines
an average sampling rate $\overline{f_{s}}=N/T$ which will lead to
$M\approx T\overline{f_{s}}/2$ as a lower limit. The parameter $T$
scales the total sampling (e.\,g. time) range interval in one dimension.
However, the question about the possible maximum frequency which can
be detected in randomly sampled data, still remains. If we assume
sampling points originating from a regular grid, but with randomly
distributed missing values, then $f_{s}\approx\min(\Delta t)^{-1}$
relates to the minimal distance between two neighboring points as
upper limit of $f_{\mathrm{s}}$. Data in such a grid is taken at
$t_{i}=t_{1}+n_{i}p$ instances, where $p$ is a kind of a common
divisor, refer to \citet{eyer1999variable} and $n_{i}$ is a non
complete set of values to reach every location. Given $n_{i}\in\mathbb{N}^{+}$
we will find that the effective maximum frequency fulfills $f_{s}>\overline{f_{s}}$
. Care must be taken with this assumption, because it could lead to
undesired large values of $f_{s}$ and therefore wrong estimations
of $M$.

\citet{horne1986aprescription} carried out an extensive study about
the number of independent frequencies (and the maximum detectable
frequency). They found an empirical approximation
\begin{equation}
M=-6.362+1.193N+0.00098N^{2},\label{eq:Ind_Freq}
\end{equation}
which is a compromise between the conservative $N/2$ and the artificially
large minimal distance value. A detailed discussion on FAP and the
independent frequencies can be found in the studies by \citet{baluev2008assessing,baluev2013detecting,baluev2013detecting2}.

\section{The multivariate Lomb-Scargle method}

\label{sec:The-multivariate-Lomb-Scargle}A signal $S$ depending
of $n$-independent variables represents a function $\mathbb{R}^{m}\rightarrow\mathbb{R}$
with the input vector described by $\vec{t}=[t_{1},t_{2},\dots,t_{m}]$.
The model function for multivariate LSM is gained by replacing the
arguments of cosine and sine in the univariate model function in Equation~(\ref{eq:LS-Model})
by vectors resulting in
\begin{equation}
Y(\vec{t})=\sum_{k=0}^{M}\left(a_{k}\cos\left(\vec{\omega}_{k}\cdot\left(\vec{t}-\vec{\tau}_{k}\right)\right)+b_{k}\sin\left(\vec{\omega}_{k}\cdot\left(\vec{t}-\vec{\tau}_{k}\right)\right)\right).\label{eq:LS_Modelfunction}
\end{equation}
In this case, the shifting parameter $\vec{\tau}_{k}\in$ $\mathbb{R^{\mathrm{\mathit{m}}}},0\leq k\leq M,$
is a vector and in principle hard to calculate. However, if the argument
of the cosine is expanded, it is obvious that the scalar product $\vec{\omega}_{k}\cdot\vec{\tau}_{k}\in\mathbb{R}$
does not depend on time and thus, the cosine argument can be written
as $\vec{\omega}_{k}\cdot\vec{t}-\tau_{k}^{*}$ with $\tau_{k}^{*}=\vec{\omega}_{k}\cdot\vec{\tau}_{k}$.
The determination of $\tau_{k}^{*}$ is similar as for $\tau$ shown
in Equation~(\ref{eq:LS-ModelTauCont}), but with some differences
in the equations since the phase, instead of the time coordinate,
is shifted now. This is shown in the following.

\subsection{Derivation of the shifting parameter}

\label{subsec:Derivation-of-tau}In this section it is shown that
shifting the phase, instead of time, does not affect the Lomb-Scargle
algorithm.The derivation of the shifting parameter for the multivariate
case is delineated for the time discrete signal $\hat{S}$$=\Bigl\{(\hat{S}_{i},\vec{t}_{i})\in R^{\mathrm{\mathit{m+1}}},0\leq i\leq N-1\Bigl\}.$
Starting with the orthogonality condition 
\begin{equation}
\sum_{n}\sin\left(\vec{\omega}_{k}\cdot\vec{t}_{n}-\tau_{k}^{*}\right)\cos\left(\vec{\omega}_{k}\cdot\vec{t}_{n}-\tau_{k}^{*}\right)=0\label{eq:LS_tau_condition}
\end{equation}
and applying the trigonometric identities to remove the differences
in the arguments:
\begin{align}
\sum_{n}\left[\left(\underset{\mathrm{cs}}{\underbrace{\cos\left(\vec{\omega}_{k}\cdot\vec{t}_{n}\right)}}\underset{\mathrm{ct}}{\underbrace{\cos\left(\tau_{k}^{*}\right)}}+\underset{\mathrm{ss}}{\underbrace{\sin\left(\vec{\omega}_{k}\cdot\vec{t}_{n}\right)}}\underset{\mathrm{st}}{\underbrace{\sin\left(\tau_{k}^{*}\right)}}\right)\right.\\
\left.\left(\underset{\mathrm{ss}}{\underbrace{\sin\left(\vec{\omega}_{k}\cdot\vec{t}_{n}\right)}}\underset{\mathrm{ct}}{\underbrace{\cos\left(\tau_{k}^{*}\right)}}-\underset{\mathrm{cs}}{\underbrace{\cos\left(\vec{\omega}_{k}\cdot\vec{t}_{n}\right)}}\underset{\mathrm{st}}{\underbrace{\sin\left(\tau_{k}^{*}\right)}}\right)\vphantom{\left(\underset{A}{\underbrace{\left(_{n}\right)}}\right)}\right] & =0\nonumber 
\end{align}
By using the defined abbreviations this equation can be simplified
to 
\[
\sum_{n}\left(\mathrm{cs}\thinspace\mathrm{ss}\thinspace\mathrm{ct^{2}}+\mathrm{ss^{2}}\thinspace\mathrm{st}\thinspace\mathrm{ct}-\mathrm{cs^{2}}\thinspace\mathrm{ct}\thinspace\mathrm{st}-\mathrm{ss}\thinspace\mathrm{cs}\thinspace\mathrm{st^{2}}\right)=0.
\]
After rearranging the summation (the terms $\mathrm{ct}$ and $\mathrm{st}$
do not depend on $n$) the following equation is deduced: 
\begin{align}
\sum_{n}(\mathrm{cs^{2}\thinspace ct\thinspace st-ct\thinspace st\thinspace ss^{2}}) & =\sum_{n}(\mathrm{cs\thinspace ss\thinspace ct^{2}-ss\thinspace cs\thinspace st^{2})}\nonumber \\
\sum_{n}\mathrm{ct\thinspace st\thinspace(cs^{2}-ss^{2})} & =\sum_{n}\mathrm{cs\thinspace ss\thinspace(ct^{2}-st^{2})}\nonumber \\
\mathrm{\frac{ct\thinspace st}{ct^{2}-st^{2}}} & =\frac{\sum_{n}\mathrm{cs\thinspace ss}}{\sum_{n}(\mathrm{cs^{2}-ss^{2}})}
\end{align}
The fraction on both sides can be simplified by applying $\cos^{2}(x)-\sin^{2}(x)=\cos(2x)$
and $\cos(x)\sin(x)=\frac{1}{2}\sin(2x)$ resulting in 
\begin{align*}
\frac{\frac{1}{2}\sin(2\tau^{*})}{\cos(2\tau^{*})} & =\frac{\frac{1}{2}\sum_{n=0}^{N-1}\sin(2\vec{\omega}_{k}\cdot\vec{t}_{n})}{\sum_{n=0}^{N-1}\cos(2\vec{\omega}_{k}\cdot\vec{t}_{n})},
\end{align*}
and then the fraction on the left hand side is replaced by the tangent
\begin{align}
\tan(2\tau^{*}) & =\frac{\sum_{n=0}^{N-1}\sin(2\vec{\omega}_{k}\cdot\vec{t}_{n})}{\sum_{n=0}^{N-1}\cos(2\vec{\omega}_{k}\cdot\vec{t}_{n})}.\label{eq:LS_tau}
\end{align}
In comparison with one dimensional LSM, the frequency $\omega_{k}$
is missing on the left side in Equation~(\ref{eq:LS_tau}).

\subsection{Parameter estimation}

\label{subsec:Lombs-Idea}Similar to the procedure for the LSM for
a discrete signal in one dimension (see Equation (\ref{eq:LSM1DStart})),
a discrete multivariate signal $(\hat{S}_{i},\vec{t}_{i})$ with $N$
samples is multiplied by $\cos(\vec{\omega}_{k}\cdot\vec{t_{i}}-\tau_{k}^{*})$
resulting in 
\begin{multline*}
\sum_{n}\hat{S}(\vec{t}_{n})\cos(\vec{\omega}_{k}\cdot\vec{t}_{n}-\tau_{k}^{*})=\sum_{n=0}^{N-1}\Bigl(a_{k}\cos^{2}(\vec{\omega}_{k}\cdot\vec{t}_{n}-\tau_{k}^{*})\\
+\underset{=0,\text{if orthogonal}}{\underbrace{b_{k}\sin(\vec{\omega}_{k}\cdot\vec{t}_{n}-\tau_{k}^{*})\cos(\vec{\omega}_{k}\cdot\vec{t}_{n}-\tau_{k}^{*})}}\Bigl).
\end{multline*}
The determination of the parameter $a_{k}$ is similar to the one-dimensional
case shown in Equation (\ref{eq:LS_Basic_ak}):
\begin{align}
\frac{\sum_{n=0}^{N-1}y(\vec{t}_{n})\cos(\vec{\omega}_{k}\cdot\vec{t}_{n}-\tau_{k}^{*})}{\sum_{n=0}^{N-1}\cos^{2}(\vec{\omega}_{k}\cdot\vec{t}_{n}-\tau_{k}^{*})} & =a_{k}.\label{eq:LS_A}
\end{align}
The parameter $b_{k}$ is calculated analogously:
\begin{align}
\frac{\sum_{n=0}^{N-1}y(\vec{t}_{n})\sin(\vec{\omega}_{k}\cdot\vec{t}_{n}-\tau_{k}^{*})}{\sum_{n=0}^{N-1}\sin^{2}(\vec{\omega}_{k}\cdot\vec{t}_{n}-\tau_{k}^{*})} & =b_{k}.\label{eq:LS_B}
\end{align}
The power spectral density (Eq.~(\ref{eq:normPSD})) as well as the
false alarm probability (Eq.~(\ref{eq:FAP})) are calculated in the
same way compared to the method in section \ref{subsec:Power-Spectral-Density}.

\section{Application}

\label{sec:Application}
\begin{figure*}
\subfloat[Recorded data with missing values]{\input{Pictures/Example_SimpWave.tex}

}\hfill{}\subfloat[\label{fig:PSD-LSM}Standardized PSD with LSM]{\input{Pictures/Example_SimpWave_Spec.tex}

}\hfill{}\subfloat[\label{fig:PSD-FT}Standardized PSD with DFT]{\input{Pictures/Example_SimpWave_SpecFFT.tex}

}

\caption{\label{fig:Simple-Wave} Comparison of the power spectral density
PSD calculated with the LSM and with the DFT approach for a two-dimensional
input data with missing values. (a) Sampled data $z=\cos(2\pi(xf_{x}+yf_{y})+\pi/4)$
for $x,y\in[-1,1]$ with $\delta x,\delta y=0.025$ and $f_{x}=3.25$
and $f_{y}=6.32$. Gray areas represent missing values. (b) PSD calculated
with the LSM and (c) PSD calculated with DFT. Notice that the systematic
error $\epsilon_{\mathrm{T}}$ and the sparse resolution leads to
a rough approximation of the frequency and to a reduced amplitude
(by 50\%) in the PSD calculated by DFT.}
\end{figure*}

The applications are ordered by the complexity of the sampling. Starting
with a regular grid with missing values for the synthetic test data,
the measurements of the ultrasound Doppler velocimetry contains jitter
and missing values. The most general scenario is then given by the
astrophysical 2D data set of sunspots, which appear freely in time
and space.

An implementation of the multivariate Lomb-Scargle method is available
in the \texttt{spectral} package published on CRAN \citep{seilmayer2019commonmethods}
to give the user access to a multi-dimensional analysis with an easy
to use interface.

\subsection{Synthetic Test Data}

Let's start with a simple test case, where the input signal is a simple
two dimensional plain wave
\[
z=\cos\left(2\pi(f_{x}x+f_{y}y)+\frac{\pi}{4}\right)
\]
with $f_{x}=3.25$ and $f_{y}=6.32$ as the dimensionless frequencies,
which are specifically selected in such a way that $f_{x},f_{y}\neq k\cdot\omega_{0}$.
Furthermore, data gaps are modeled by removing randomly distributed
and uncorrelated grid points covering 60\% of the total data set.
As LSM requires an appropriate input vector of frequencies we choose
for both variables $f_{x},f_{y}\in[-10,10]$ with a resolution of
$\delta f_{x},\delta f_{y}=0.025$. Figure~\ref{fig:Simple-Wave}a
shows the nonuniform distributed input data in the rage of $x,y\in[-1,1]$
with $\delta x,\delta y=0.025$. With respect to the chosen frequencies
$f_{x}$ and $f_{y}$ it becomes evident, that both frequency parts
do not fit to the data range by an integer fraction. Gray areas indicate
missing (non available numbers ``$\mathrm{nan}$'') values.

The corresponding power spectral density shown in Figure~\ref{fig:PSD-LSM}
displays the maxima at the 1st and 3rd quadrant, which represents
a wave traveling upwards. The figure illustrates that even with a
huge amount of missing data it is possible to properly detect the
periodic signal. In the present case the value of $\mathrm{psd}(f_{x},f_{y})\approx1$
indicates a perfect fit to the corresponding sinusoidal model. For
comparison Figure~\ref{fig:PSD-FT} depicts the standardized PSD
(compare with Eq.~\ref{eq:normPSD})
\begin{equation}
\mathrm{psd_{FT}}(\vec{\omega})=\frac{N}{N-1}\cdot\frac{1}{2\sigma_{0}^{2}}\left(\frac{2\left|\mathcal{F}\left(z(x,y)\right)\right|}{1-\frac{N_{\mathrm{Zero}}}{N}}\right)^{2}\label{eq:PSD_FT}
\end{equation}
calculated from a discrete Fourier transform symbolized by the operator
$\mathcal{F}\bigl(g(x)\bigl)(\omega)=\int g(x)\e^{\im\omega x}\diff x$.
Here missing values ($\mathrm{nan}$) are handled by zero padding
which means that the introduced gaps are filled with $0$. Because
Fourier transform represents a conservative transformation (mapping)
into the spectral domain, zero padding lead to a significant reduction
of the average signal energy. A proper rescaling to the number of
zeros $N_{\mathrm{Zero}}$ (missing values) is required to ensure
approximated amplitude estimations. Zero padding always changes the
character of the input signal in a way that the corresponding DFT
threats the zeros as if they where part of the original signal. The
resulting PSD becomes finally different from the original one.

Since the signal frequencies $f_{x},f_{y}\neq k/T_{x,y}$ do not fit
into an integer spaced scheme ($k\in\mathbb{N}$) of the data range
$T_{x,y}=$ 2, the discrete Fourier transform suffers from its drawbacks,
which where briefly discussed before. The effect of leakage and misfit
of frequencies can be recognized in Figure~\ref{fig:PSD-FT} in terms
of a rather coarse resolution and a lower PSD value.

\subsection{3D UDV Flow Measurement}

\begin{figure*}
\begin{centering}
\input{Pictures/SensorData.tex}
\par\end{centering}
\caption{\label{fig:MRI_Data}Sensor data in the co-rotating reference frame.
Data is taken as time series from two rotating sensors. The measurement
time of the sensor data is folded with its individual phase location,
so $\varphi_{\mathrm{S1}}=\omega t$ and $\varphi_{\mathrm{S2}}=\omega t+\pi$
describe the two ordinates.}
\end{figure*}
The second example is taken from previous experimental flow measurements
from \citet{seilmayer2014experimental} for the magneto rotational
instability \citet{seilmayer2014experimental}. The experiment consists
of a cylindrical annulus containing a liquid metal between the inner
and the outer wall. In this experiment the inner wall is rotating
at a frequency $\omega_{\mathrm{in}}=2\pi\cdot\unit[0.05]{Hz}$ and
the outer wall at a frequency $\omega_{\mathrm{out}}=2\pi\cdot\unit[0.013]{Hz}$.
The flow is driven by shear since $\omega_{\mathrm{in}}-\omega_{\mathrm{out}}>0$.
Two ultrasound sensors mounted at the outer cylinder, on opposite
side of each other, measure the axial velocity component $v_{\mathrm{z}}$
along the line of sight parallel to the rotation axis of the cylinder.
Additionally, the liquid metal flow is exposed to a magnetic field
$B_{\varphi}\propto r^{-1}$ originating from a current $I_{\mathrm{axis}}$
on the axis of the cylinder. A typical time series of the axial velocity
(along the measuring line) is displayed in the Figure~\ref{fig:MRI_Data}.
The existence of periodic patterns in this figure indicates a traveling
wave propagating in the fluid. In cylindrical geometry the traveling
wave can be expressed as
\begin{equation}
v_{\mathrm{z}}(t,r,z,\varphi)\propto v_{0}(r)\e^{\im(\omega t+kz+m\varphi)},
\end{equation}
with $\omega=2\pi f$ as the corresponding drift frequency, $k$ as
the vertical spatial structure and $m$ as the azimuthal symmetry.
A closer look to the picture indicates that this wave is not axial
symmetric with $m=1$ which is proven with data analysis in the following.

The preparation of data requires a mapping 
\begin{align}
v_{\mathrm{z}} & :=f(t_{n},d_{n},\varphi_{1,n},\varphi_{2,n})
\end{align}
with
\begin{equation}
\varphi_{1,n}=\omega_{\mathrm{out}}t_{n}\:\text{(sensor 1, S1)}
\end{equation}
and 
\begin{equation}
\varphi_{2,n}=\omega_{\mathrm{out}}t_{n}+\pi\thinspace\text{(sensor 2, S2)}
\end{equation}
with the azimuthal angles $\varphi_{2,n}=\varphi_{1,n}+\pi$ depending
on time as the sensors are attached to the outer wall. The subscript
$n$ indicates the sampling in this sense. Finally, the measured time
series $v_{z}$ depends on time $t_{n}$, depth $d_{n}$ and angular
position $\varphi_{n}$. This fits perfectly to the proposed multivariate
LSM.

Figure~\ref{fig:MRI_Spektrum} illustrates the result of the LSM
decomposition into several $m$-modes showing the amplitudes. The
$m=0$ mode contains a stationary structure at $f\approx0$, which
originates from sensor miss-alignments and (thermal) side effects
in the flow. The minor non-stationary components, here the two point-symmetric
peaks, originate from cross-talk (or alias projections) of the $m=1$
mode for two reasons: (i) because the sensors behave not exactly identical
(e.\,g. misalignment or different sensitivity) and therefore respond
slightly different to the same flow signal. This means that one of
the sensors projects a little bit more energy into the data than the
other. This leads to a ``leakage''-effect and a weak signal in $m=0$
mode. (ii) the outer rotation of the sensors acts like an additional
sampling frequency $f_{\mathrm{out}}$. In consequence, the spectrum
of this regular sampling function folds with the spectrum of the observed
process leading to alias images (copies) of the original process spectrum
into other frequency ranges (i.\,e. $m$'s). That is why the non
stationary and point-symmetric (with reference to the origin) signals
in the $m=0$ panel correspond to the (mirrored) patterns originally
present in the $m=1$ panel.

The AMRI wave itself is located in the $m=1$ panel with a characteristic
frequency in time ($f$) and space ($k$). Here, two components can
be identified: the dominant wave at $f\approx\unit[9]{mHz}$ and $k\approx\unit[20]{m^{-1}}$
and a minor counterpart at $f\approx\unit[4]{mHz}$ and $k\approx\unit[-20]{m^{-1}}$.
The signals in $m=3$ and $m=5$ might be assigned to aliases from
the inner rotation $f_{\mathrm{i}}=\unit[0.05]{Hz}$.

\begin{figure*}
\input{Pictures/SensorSpectrum_mod.tex}

\caption{\label{fig:MRI_Spektrum}Spectrum of sensor data. The amplitude spectrum
is calculated from the data in Fig.~\ref{fig:MRI_Data} with LSM.
The weak alias peaks in $m=0$ panel originate from sensor mismatch
and aliasing effect due to the outer rotation. The latter acts like
a sampling frequency and therefore projects the AMRI wave to $m=0$.}
\end{figure*}

The advantage of ``high'' dimensional spectral decomposition is
the improved noise rejection. With respect to the raw data given in
Figure~\ref{fig:MRI_Data} it is obvious that high frequency noise
is present in the data. Depending on the exact distribution of noise
its energies spread over a certain range of frequencies. If we would
select a representative depth and angle so that $d_{n},\varphi_{n}=\text{const.}$,
the velocity $v_{z}$ only depends on time. In the subsequent one
dimensional analysis the noise would accumulate along the single frequency
ordinate probably hiding the signal of interest. Taking the higher
order analysis distributes the noise energies over multiple domain
variables. For the present example this means that the distortions
are projected into higher frequencies $f$, higher $m$'s and larger
$k$'s. Since the signal of interest remains in the same spectral
corridor, its signal amplitude becomes more clear, because of the
``reduced'' local noise.

\subsection{\label{subsec:Analyzing-2D-sun}Analyzing 2D sunspot data}

\begin{figure*}
\centering{}\input{Pictures/SunSpot_Butterfly.tex}\caption{\label{fig:SunSpotButterfly-diagram}Butterfly diagram of the sunspot
data. The separation of the individual wings took place according
to the proposed procedure presented in \citet{leussu2016properties}.
The colored patches indicate the changing polarity of each cycle.
The tilted segmentation lines depict the optimized borders between
two cycles. }
\end{figure*}
 Since the beginning of the 17th century systematic visual observations
of sun spots are available which allow to investigate dynamic processes
taking place in the sun. The appearance of sunspots on the surface
of the Sun depends on the level of solar activity and therefore it
renders some fundamental features of the underlying solar dynamo,
such as the $\unit[11]{yrs}$ solar cycle. Since the beginning of
the 1820's observational data is available in terms of a two dimensional
time series that exhibits a periodic wing-like pattern essentially
symmetric with respect to the Sun's equator. The sunspot butterfly
diagram emerges and summarizes the individual sunspot groups appearing
at a certain time and latitude on the Sun for the past 190 years.
Additionally, Figure~\ref{fig:SunSpotButterfly-diagram} depicts
the assigned field polarity order indicated by color (gray or black).
The segmentation procedure followed the suggestions from \citet{leussu2016properties},
but with some simplifications leading only to minor miss-assignments
for individual sun spot groups. The field polarity $P(Y,L)\in\{-1,1\}$
gives the arrangement of leading North/South polarity of a sunspot
group, depending on the year $Y$ and the mean latitude $L$. This
polarity is changing approximately each 11 years which is related
to Schwabe's cycle. The full period of $\unit[22]{yrs}$ is then called
the Hale cycle.

The data is taken \textquotedbl as-is\textquotedbl{} from \citet{leussu2017wingsof},
which originates from several sources with different qualities, i.~e.
the Royal Greenwich Observatory \textendash{} USAF/\-NOAA(SOON)\footnote{Available at\\
 \url{http://solarscience.msfc.nasa.gov/greenwch.shtml}}\-\citep{clette2014revisiting,willis2016reexamination}, Schwabe\footnote{Available at\\
 \url{http://www.aip.de/Members/rarlt/sunspots/schwabe}}\-\citep{arlt2013sunspot} and Spoerer\footnote{the reader might also refer to the historic publications \citet{spoerer1889memoires,spoerer1890profspoerers}}\-\citep{diercke2015digitization}
data sets. A detailed discussion of the data collection is given by
\citet{leussu2016properties,leussu2017wingsof} and the referenced
literature.

The following example deduces the spectral decomposition from this
unevenly sampled binary data set $P(Y,L)$ to identify typical periods
and their modulation. In contrast to the two examples above, the sunspot
data set consists of real arbitrary sampling in both, time and space,
giving the most general scenario for the LSM.

The starting point for the subsequent analysis is the simplified segmentation
of the butterfly diagram reassigning the field polarity. The segmentation
takes place by optimizing the distance $d$ of each point $P(L,Y)$
with respect to the segmentation line
\[
L(Y)=\begin{cases}
m_{1}\left(Y-T_{0}\right) & m_{1}<0,L>0\\
m_{2}\left(Y-T_{0}\right) & m_{2}>0,L\leq0
\end{cases},
\]
which is a piece wise linear function with the intersection point
$\{T_{0},0\}$. The dependent variables $Y$ and $L$ describe the
time in years and latitude in degrees respectively. The individual
slopes $m_{1},m_{2}<\unit[12]{deg/yrs}$ are assigned on the northern
and southern hemispheres, respectively. The initial value of the temporal
shift $T_{0g}\approx1835+11g$ considers the group id $g$ to move
forward in the data set. The optimization then minimizes the penalty
function $p\sim\e^{-d}(1+m/5)$ as a measure of inverse distance between
points and corresponding segmentation lines and minimal slope. 
\begin{figure}
\begin{centering}
\input{Pictures/SunSpot_singleGroup.tex}
\par\end{centering}
\caption{\label{fig:SegmentationLines}Optimization of segmentation lines.
Black dots represent the sunspot positions over time. The gray line
indicate the initial condition for the fit whereas the red one depicts
the result of the optimization.}
\end{figure}
 Figure~\ref{fig:SegmentationLines} depicts the optimization group
of the last $\unit[30]{yrs}$, where the gray line indicates the initial
condition and red line the optimized result. The subdivision is straight
forward but not perfect, indicated for example by several points around
the year 1990 at $\pm50^{\circ}$ latitude which might be assigned
differently when doing the suggested segmentation by \citet[Fig. 2]{leussu2017wingsof}.
However, with a total sum of $\mathcal{O}(2.8\cdot10^{5})$ observed
sunspots, it is unlikely that such single distortions change the character
of the spectrum. The final segmentation is depicted in Figure~\ref{fig:SunSpotButterfly-diagram}
where 18 individual cycles subdivided by a \textquotedbl\textgreater\textquotedbl{}
- shaped separation area are shown.

\begin{table}
\caption{\label{tab:Parameters-of-Segmentation}Parameters of Segmentation
Lines. The slopes $m_{1,2}$ are given in $\mathrm{deg/year}$}

\centering{}%
\begin{tabular}{rrrrrrr}
\toprule 
$T_{0}/\mathrm{Year}$ & $m_{1}$ & $m_{2}$ &  & $T_{0}/\mathrm{Year}$ & $m_{1}$ & $m_{2}$\tabularnewline
\midrule 
1835.1 & -11.8 & 13.2 &  & 1946.3 & -8.9 & 5.9\tabularnewline
1846.0 & -8.5 & 8.5 &  & 1956.9 & -7.2 & 6.4\tabularnewline
1858.4 & -7.9 & 8.7 &  & 1967.3 & -6.1 & 7.4\tabularnewline
1868.6 & -12.6 & 11.4 &  & 1979.7 & -4.2 & 5.0\tabularnewline
1880.3 & -5.5 & 10.0 &  & 1989.1 & -7.3 & 7.3\tabularnewline
\addlinespace
1892.0 & -5.6 & 6.2 &  & 1999.7 & -5.7 & 5.7\tabularnewline
1902.7 & -6.1 & 8.1 &  & 2011.3 & -6.5 & 6.7\tabularnewline
1914.0 & -6.1 & 20.0 &  &  &  & \tabularnewline
1925.0 & -9.1 & 9.0 &  &  &  & \tabularnewline
1935.4 & -11.7 & 12.3 &  &  &  & \tabularnewline
\midrule 
Average $\overline{m}_{1,2}$ & -7.68 & 8.94 &  &  &  & \tabularnewline
\bottomrule
\end{tabular}
\end{table}
\begin{figure*}
\begin{centering}
\input{Pictures/SunSpot_Spectrum.tex}
\par\end{centering}
\caption{\label{fig:SunSpotSpectrum}2D LSM spectrum of the sunspot data. The
numbering of the individual maxima correspond to the ID in Table~\ref{tab:Significant-peaks}.
The curved lines follow a path of constant phase velocity with $v=(f_{\mathrm{Lat}}P_{\mathrm{Yrs}})^{-1}=\mathrm{const.}$
indicating a wave structure with a certain time dependence. The bold
black lines depicts the mean waves with $v=\text{\ensuremath{\overline{m}_{1,2}}}$,
whereas the dashed lines follow a perfect sinusoidal wave defined
by the dominant points $\mathit{ID}\in\{32,33\}$. The range of the
solar cycle period is given by the vertical dashed gray lines.}
\end{figure*}

By observing the slopes of segmentation lines and the individual shapes
of the patches a slight asymmetry between northern and southern hemisphere
can be recognized. This becomes clear in the different average values
of the slopes (see Table~\ref{tab:Parameters-of-Segmentation}).
Figure~\ref{fig:SunSpotButterfly-diagram} also indicates a modulation
of the maximum latitude of individual cycles with a period of $\sim$200
years which may be related to the Suess de Vries cycle.

The spectral LSM decomposition is based on the definition of a two
dimensional wave model
\begin{align}
P(Y,L)\sim & A\cdot\cos(\omega_{\mathrm{Yrs}}\mathit{Y}+\omega_{\mathrm{Lat}}\mathit{L}+\tau^{*})+\label{eq:WaveModel}\\
 & B\cdot\sin(\omega_{\mathrm{Yrs}}\mathit{Y}+\omega_{\mathrm{Lat}}\mathit{L}+\tau^{*})+\epsilon,\nonumber 
\end{align}
where $P$ is given by the assigned patch polarity. To show the advantage
and robustness of LSM the raw data is supplied without any further
preprocessing. The selected frequencies $\omega_{\mathrm{Yrs},n},\omega_{\mathrm{Lat},n}\propto n^{-1}$
are given on a rectangular grid with inverse distance so that the
periods $\omega^{-1}$ are distributed uniformly alongside the consecutive
counter $n$. Furthermore, the frequency resolution in the range of
$|\omega_{\mathrm{Lat}}|=0$ is selected finer to better resolve this
region.

Figure~\ref{fig:SunSpotSpectrum} gives the resulting spectral decomposition
in a reciprocal $\log$-scale plot to focus on the periods. Each of
the selected peaks (numbered black dots) provides a FAP value of $p<10^{-10}$,
meaning that these periods are significantly different from noise
level. Each of these points originate from a maximization of the local
amplitude value as a subsequent refinement. The numbering corresponds
to the peak identifier $\mathit{ID}$ in Table~\ref{tab:Significant-peaks}.
The binary order information $\{-1,1\}$ introduces higher harmonics
into the spectrum, which do also appear with a significantly low FAP
value. Table~\ref{tab:Significant-peaks} summarizes peaks with $P_{\mathrm{Yrs}}>\unit[3]{yrs}$
and a certain strength. A detailed discussion about the definition
of the noise level in case of binary input data and the proper peak
selection is left for future work.

The interpretation of the spectrum can be given in two ways. First,
a purely temporal frequency analysis provides good agreement between
the peaks found and the common known periods. Table~\ref{tab:SunSpot_Common-Peaks}
summarizes the major outcome in comparison with literature. For example
the $\unit[22]{yrs}$ Hale cycle varies from $\unit[18\dots28]{yrs}$
(\citet{usoskin2017ahistory}) which is identified by the two main
peaks $\mathit{ID}\in\{32,33\}$. Since the solar cycle is modulated,
refer to \citet{hathaway2015thesolar}, it is natural that a broad
spectrum with many harmonics becomes present. These subsequent patterns
are related to a set of local maxima mainly collapsing on a horizontal
line at $\left|f_{\mathrm{Lat}}\right|\approx\unit[1.4\cdot10^{-2}]{deg^{-1}}$.
The interpretation is, that the complex structure of the individual
wing-shapes of the butterfly diagram with different widths, heights
and orientation, correspond to a main period $P_{\mathrm{Hale}}=\unit[21.634]{yrs}$
and some harmonics. Next to that we can identify typical periods which
are related to the Gleissberg process (see Table~\ref{tab:SunSpot_Common-Peaks}).
The short periods in the range of $P\approx\unit[7.2]{yrs}$ are consistent
with the data provided by \citet{prestes2006spectral}, \citet{kane1997quasibiennial}
and partly with \citet{deng2020periodic}.
\begin{table*}
\caption{\label{tab:SunSpot_Common-Peaks}Common Peaks. Period values are given
in years. Values above $\unit[100]{yrs}$ are affected by the low
period resolution caused by the limited time span of sunspot data.
The results from \citet{prestes2006spectral} refer to Schwabe cycle
and therefore are doubled.}

\begin{centering}
\begin{tabular}{>{\centering}p{2.5cm}lll}
\toprule 
Process & Common Period & From spectrum & Ref.\tabularnewline
\midrule
Eddy & 515 & 559 & 1\tabularnewline
 & 350 & 359 & 1\tabularnewline
Hale (Schwabe) & 22.14 (11.07) & 21.63 $\pm$ 2.5 & 2\tabularnewline
 & 18...28 (2) &  & \tabularnewline
Gleissberg & 88 (80...150) & 78, 85, 125, 156 & 1,2\tabularnewline
\textendash{} & 126 & 125 & 3\tabularnewline
\textendash{} & 2$\times$3.6 & 7.2 & 4\tabularnewline
\textendash{} & 2$\times$3.9 & 7.7 & 4,5\tabularnewline
\bottomrule
\end{tabular}
\par\end{centering}
(1) \citet{mccracken2013theheliosphere}; (2) \citet{usoskin2017ahistory};
(3) \citet{ogurtsov2002longperiod}; (4) \citet[Fig. 6]{prestes2006spectral};
(5) \citet{kolotkov2015hilberttextendashhuang}.
\end{table*}

Second, the latitudinal dimension of the spectrum spreads the peaks
vertically. This is a great advantage compared to a purely 1D analysis,
where all signal would be projected on the $f_{\mathrm{Lat}}=0$ line.
In this case individual peaks, i.\,e. $\mathit{ID}\in\{17,21,25,34,27,\dots\}$
would collapse on the ordinate in a single domain analysis. Furthermore,
Figure~\ref{fig:SunSpotButterfly-diagram} can be interpreted as
a set of modulated waves propagating towards the equator. The two
dimensional LSM decomposes the frequencies and gives access to the
properties of such waves.

If a single large peak corresponding e.\,g. to $P\approx\unit[22]{yrs}$
and with $|f_{{\rm Lat}}|>0$ would be related to a perfect sinusoidal
wave, the associated complex texture of moving patterns is neglected.
In contrast to that, a constant pattern, i.\,e. a fixed group of
sunspots drifting towards the equator, would find its representation
in a variety of peaks along a line of constant phase velocity, $v=(f_{\mathrm{Lat}}P_{\mathrm{Yrs}})^{-1}=\mathrm{const.}$.
If such a pattern slowly changes over time, which is clearly the case
in Figure~\ref{fig:SunSpotButterfly-diagram}, the corresponding
$v$-line becomes the center frequency of an amplitude modulated wave.
Hereby, the average phase velocity, represented by the $v$-line,
keeps the same but is accompanied by side bands to the left and to
the right. This can be seen as an $n$D amplitude modulation, where
the $v$-line is the carrier frequency.

Figure~\ref{fig:SunSpotSpectrum} provides some of the $v$-lines
related to typical referenced processes. Starting with the Hale-cycle,
which is identified by $P_{\mathrm{Hale}}=\unit[21.634]{yrs}$ and
$f_{\mathrm{Lat}}=\unit[1.152\cdot10^{-2}]{deg^{-1}}$ ($\mathit{ID}\in\{32,33\}$),
we obtain the typical phase velocity of $v_{\mathrm{Hale}}=\unit[4.012]{deg/yrs}$.
This single phase approximation assumes a perfect sinusoidal wave
represented by a single peak. Nevertheless, the dashed black line
indicates the corresponding $v$-line which includes the main peak
of the diagram but passes many others. On the other hand, the average
slopes of the segmentation lines for the northern and southern hemisphere
can be deduced from Figure~\ref{fig:SunSpotButterfly-diagram}. Here
the mean values, compared with Table~\ref{tab:Parameters-of-Segmentation}
result in the mean slope velocities $\overline{v}_{1,2}=\overline{m}_{1,2}$
which correspond to the black solid lines. As the butterfly diagram
indicates a time varying pattern (of the individual wings) it is quite
natural, that the mean phase velocity is surrounded by side bands.
The concept of side band modulation assigns such peaks to the same
a complex wave pattern, like the Hale-wave. Hereby the (carrier) $v$-lines
touch certain local maxima, i.\,e. at $P\approx4$ or $\mathit{ID}\in\{57\dots62\}$,
so they can be assumed to map the most realistic wave structure for
the butterfly diagram in contrast to a purely sinusoidal Hale - wave
pointed out above. The argument about the mapping of certain points
to a characteristic $v$-line is supported by the fact that in this
case $\overline{v}$ originates from the space-time analysis of Figure~\ref{fig:SunSpotButterfly-diagram}.
The conclusion is that the Hale-cycle, finding its main period at
$P_{\mathrm{\ensuremath{Hale}}}$ might be described as a modulated
set of waves with minor prominent harmonics. If so, peaks arranged
around the characteristic $v$-line such as $\mathit{ID}\in\{51\dots54\}$
or $\mathit{ID}\in\{48,49\}$ , are directly coupled with the Sun
cycle. These individual side bands $\mathit{ID}\in\{51\dots54\}$
are covered by the results of \citet{prestes2006spectral} investigating
geomagnetic indices and sunspot number time series. The peaks $\mathit{ID}\in\{48,49\}$,
are supported by findings from \citet{kolotkov2015hilberttextendashhuang}
analyzing $\unit[10.7]{cm}$ radio flux measurements, helioseismic
frequency shift and the sunspot area. Short term oscillations ($\mathit{ID}\in\{17,22,23,53\}$
as doubled period) are obtained by \citet{deng2020periodic} from
grouped solar flare as well as sun spot number time series.

Despite the modulated structure of the wings, it might be of interest
to focus on their mean path from which a time dependent drift velocity
can be deduced. From the standard law for this equatorial drift, see
\citet[Eq. (4)]{hathaway2011astandard}, the mean velocity $\overline{v}=\unit[1.8622]{deg/yrs}$
follows from the average (centroid) position, $\overline{\lambda}(t)=28^{\circ}\e^{-12t/90}$
of the sunspots as function of time $t$ in a $T=\unit[12]{yrs}$
range. The corresponding bold gray $v$-line in Figure~\ref{fig:SunSpotSpectrum}
assigns several numbered and unnumbered peaks to the motion of sunspots.
The neighboring thin solid gray line is defined by $\overline{v}=\unit[1.56]{deg/yrs}$
as suggested by \citet{li2001latitude} investigating the same process
of equatorial drift.

As a speculation about other present waves, the dashed gray line defined
by the peaks $\mathit{ID}=\{2,5,6,7,13\}$ might be related to a Gleissberg
process. The corresponding line in the lower half plane of Figure~\ref{fig:SunSpotSpectrum}
is not symmetric indicating the asymmetry between the hemispheres.
And finally a remaining $v$-line could be defined by the points $\mathit{ID}=\{4,12,17,22,24,30\}$
and $\mathit{ID}=\{3,10,16,20,21,31\}$ which might belong to another
(unidentified) process.

Taking only the raw data of the sunspots (time and latitude) into
account, LSM allows the analysis of complex dependencies between individual
spectral peaks due to the idea of a modulated traveling wave. The
result of LSM was verified with characteristic peaks obtained from
a variety of references based on complementary data sets. Since the
features of the sunspots are related to the solar dynamo the analysis
has the potential to obtain a better insight into the magnetic field
generating processes in the tachocline with respect to the differential
rotation.

\section{Conclusions}

\label{sec:Conclusions}In the present work a multidimensional extension
of the Lomb-Scargle method is developed. The key aspect is the redefinition
of phase argument to $\vec{\phi}_{\mathrm{new}}=\vec{\omega}\cdot\vec{t}-\tau^{*}$.
We suggest using a modified shifting parameter $\tau^{*}$ in contrast
to the traditional approach which shifts the ordinate $\phi_{\mathrm{orig}}=\omega\cdot\left(t-\tau\right)$
instead of the phase. This enables multivariate modeling with one
single scalar value $\tau^{*}$ for all independent variables as there
is always a shifting parameter $\tau^{*}$ for which $\int_{a}^{b}\sin(\phi(t))\cos(\phi(t))\thinspace\diff t$
vanishes on any interval $[a,b]$.

The examples from Section~\ref{sec:Application} underline the strengths
of the developed procedure in a consecutive way. First the application
on ideal two dimensional test data shows the ability to analyze fragmented
time series. Here the sampling remains regular meaning $t=nT_{0}$
with $n\in\mathbb{N}$, but with $n_{i}-n_{i+1}\neq1$ as an incomplete
set of locations to describe missing values. A second quasi similar
situation is given in the experimental data set of UDV measurements,
expect a certain jitter. The dimensionality in this example was extended
to $\mathbb{R}^{3}$ to decompose the wave parameters in frequency
$f$, symmetry $m$ and spatial frequency $k$.

As the third analysis, the sunspot data sets represent the most general
case of non-evenly sampling. Taking only the time series of positions
into account, LSM is able calculate the spectrum on a individual frequency
grid pronouncing the low frequencies close to zero. The assignment
of $\{-1,1\}$ to each data point is a minor modification so the data
can be assumed as binary raw data. The spectral decomposition with
LSM shows the characteristic footprint of waves and its dynamics present
in the butterfly diagram. The time frequency spectrum itself provides
the commonly known frequencies, i.\,e. Hale-Cycle or Gleissberg process.
The second frequency $f_{\mathrm{Lat}}$ provides information about
the minor peaks which are spread out into the latitudinal domain.
Moreover, from these values the sunspot drift motion can be calculated
using a characteristic $v$-line with its side bands. For validation
of these new procedure different measures are compared with characteristic
peaks in the spectrum: (i) the mean slopes of the segmentation lines
from Figure~\ref{fig:SunSpotButterfly-diagram} can be interpreted
as average phase velocity. The corresponding $v$-line nicely assigns
several peaks to the dominant Hale cycle. (ii) the individual values
for latitude migration from \citet{li2001latitude,hathaway2011astandard}
do form a $v$-line describing sunspot motion. In any instance the
spectral signature of a moving complex pattern becomes evident by
a center line of constant phase velocity and its side bands, containing
the temporal dynamics (modulation) of the motion.

With respect to the evaluation of measurement results, the noise rejection
$\epsilon_{\mathrm{LS}}\propto4N^{-0.5}/\pi$ and confidence intervals
($\Delta a_{k}$ and $\Delta b_{k}$) of model parameters $a_{k}$
and $b_{k}$ are delineated. To emphasize the advantages of the LSM
the traditional Fourier mode decomposition is compared. It turned
out that the systematic error $\epsilon_{\mathrm{T}}$ does not vanish
for FT based methods with increasing number of samples on a fixed
interval $T$, which finally lead to the common leakage effect as
seen in Figure~\ref{fig:PSD-FT}. Here the signal amplitudes are
distributed on an area of neighboring pixels. We conclude that the
standard orthogonal mode related procedures do not represent a consistent
estimator for model parameters $a_{k}$ and $b_{k}$. Whereas the
introduction of $\tau^{*}$ in LSM leads to a consistent estimator
with $\epsilon_{\mathrm{T}}=0$ even for higher dimensions. Finally,
it was shown that LSM converges to the true model parameters with
increasing number of samples and provides a better noise rejection
($\epsilon_{\mathrm{LS}}<\epsilon_{\mathrm{FS}}$) as well.

\section*{Acknowlegdements}

The authors like to thank Rainer Arlt from Astrophysical Institute
Potsdam providing the sunspot data and Andre Gieseke from HZDR for
many fruitful discussions. F.\,Garcia kindly acknowledges the Alexander
von Humboldt Foundation for its financial support.

\bibliographystyle{aa}
\bibliography{COL_refs}

\begin{thebibliography}{61}
\expandafter\ifx\csname natexlab\endcsname\relax\def\natexlab#1{#1}\fi

\bibitem[{Al-Ani \& Tarczynski(2012)}]{al-ani2012evaluation}
Al-Ani, M. \& Tarczynski, A. 2012, Signal Processing, 92, 2484

\bibitem[{Arlt {et~al.}(2013)Arlt, Leussu, Giese, Mursula, \&
  Usoskin}]{arlt2013sunspot}
Arlt, R., Leussu, R., Giese, N., Mursula, K., \& Usoskin, I.~G. 2013, Monthly
  Notices of the Royal Astronomical Society, 433, 3165

\bibitem[{Babu \& Stoica(2010)}]{babu2010spectral}
Babu, P. \& Stoica, P. 2010, Digital Signal Processing, 20, 359

\bibitem[{Baluev(2008)}]{baluev2008assessing}
Baluev, R.~V. 2008, Monthly Notices of the Royal Astronomical Society, 385,
  1279

\bibitem[{Baluev(2013{\natexlab{a}})}]{baluev2013detecting2}
Baluev, R.~V. 2013{\natexlab{a}}, Monthly Notices of the Royal Astronomical
  Society, 436, 807

\bibitem[{Baluev(2013{\natexlab{b}})}]{baluev2013detecting}
Baluev, R.~V. 2013{\natexlab{b}}, Astronomy and Computing, 3-4, 50

\bibitem[{Barning(1963)}]{barning1963thenumerical}
Barning, F. J.~M. 1963, Bulletin of the Astronomical Institutes of the
  Netherlands, 17, 22

\bibitem[{Cadena {et~al.}(2016)Cadena, Carlone, Carrillo, Latif, Scaramuzza,
  Neira, Reid, \& Leonard}]{cadena2016pastpresent}
Cadena, C., Carlone, L., Carrillo, H., {et~al.} 2016, IEEE Transactions on
  Robotics, 32, 1309

\bibitem[{Clette {et~al.}(2014)Clette, Svalgaard, Vaquero, \&
  Cliver}]{clette2014revisiting}
Clette, F., Svalgaard, L., Vaquero, J.~M., \& Cliver, E.~W. 2014, Space Science
  Reviews, 186, 35

\bibitem[{Cohen(1995)}]{cohen1995timefrequency}
Cohen, L. 1995, Time-Frequency Analysis, Prentice {Hall} signal processing
  series (Englewood Cliffs, N.J: Prentice Hall PTR)

\bibitem[{Cumming {et~al.}(1999)Cumming, Marcy, \& Butler}]{cumming1999thelick}
Cumming, A., Marcy, G.~W., \& Butler, R.~P. 1999, The Astrophysical Journal,
  526, 890

\bibitem[{Deng {et~al.}(2020)Deng, Mei, \& Wang}]{deng2020periodic}
Deng, H., Mei, Y., \& Wang, F. 2020, Research in Astronomy and Astrophysics,
  20, 022

\bibitem[{Diercke {et~al.}(2015)Diercke, Arlt, \&
  Denker}]{diercke2015digitization}
Diercke, A., Arlt, R., \& Denker, C. 2015, Astronomische Nachrichten, 336, 53

\bibitem[{Eyer \& Bartholdi(1999)}]{eyer1999variable}
Eyer, L. \& Bartholdi, P. 1999, Astronomy and Astrophysics Supplement Series,
  135, 1

\bibitem[{Fessler \& Sutton(2003)}]{fessler2003nonuniform}
Fessler, J. \& Sutton, B. 2003, IEEE Transactions on Signal Processing, 51, 560

\bibitem[{Fessler(2002)}]{fessler2002iterative}
Fessler, J.~A. 2002, 11

\bibitem[{Frydman {et~al.}(2003)Frydman, Lupulescu, \&
  Scherf}]{frydman2003principles}
Frydman, L., Lupulescu, A., \& Scherf, T. 2003, Journal of the American
  Chemical Society, 125, 9204, publisher: American Chemical Society

\bibitem[{Geneva {et~al.}(2018)Geneva, Eckenhoff, \&
  Huang}]{geneva2018asynchronous}
Geneva, P., Eckenhoff, K., \& Huang, G. 2018, in 2018 {IEEE} {International}
  {Conference} on {Robotics} and {Automation} ({ICRA}) (Brisbane, QLD: IEEE),
  1--6

\bibitem[{Giraudeau \& Frydman(2014)}]{giraudeau2014ultrafast}
Giraudeau, P. \& Frydman, L. 2014, Annual Review of Analytical Chemistry, 7,
  129

\bibitem[{Gonzalez \& Woods(2008)}]{Gonzalez2008}
Gonzalez, R.~C. \& Woods, R.~E. 2008, Digital Image Processing, 3rd edn.
  (Prentice Hall)

\bibitem[{Greengard \& Lee(2004)}]{greengard2004accelerating}
Greengard, L. \& Lee, J.-Y. 2004, SIAM Review, 46, 443

\bibitem[{Haacke(1999)}]{Haacke1999}
Haacke, E. 1999, Magnetic Resonance Imaging: Physical Principles And Sequence
  Design (John Wiley and Sons)

\bibitem[{Hathaway(2011)}]{hathaway2011astandard}
Hathaway, D.~H. 2011, Solar Physics, 273, 221

\bibitem[{Hathaway(2015)}]{hathaway2015thesolar}
Hathaway, D.~H. 2015, Living Reviews in Solar Physics, 12, 4

\bibitem[{Hocke(1998)}]{hocke1998phaseestimation}
Hocke, K. 1998, Annales Geophysicae, 16, 356

\bibitem[{Horne \& Baliunas(1986)}]{horne1986aprescription}
Horne, J.~H. \& Baliunas, S.~L. 1986, The Astrophysical Journal, 302, 757

\bibitem[{James(2011)}]{James2011}
James, J. 2011, A Student's Guide to Fourier Transforms: with Applications in
  Physics And Engeneering, 3rd edn. (Cambridge University Press)

\bibitem[{Jerri(1977)}]{jerri1977theshannon}
Jerri, A.~J. 1977, Proceedings of the IEEE, 65, 1565

\bibitem[{Kane(1997)}]{kane1997quasibiennial}
Kane, R.~P. 1997, Annales Geophysicae, 15, 1581

\bibitem[{Kolotkov {et~al.}(2015)Kolotkov, Broomhall, \&
  Nakariakov}]{kolotkov2015hilberttextendashhuang}
Kolotkov, D.~Y., Broomhall, A.-M., \& Nakariakov, V.~M. 2015, Monthly Notices
  of the Royal Astronomical Society, 451, 4360

\bibitem[{Kov{\'a}cs(1981)}]{kovacs1981frequency}
Kov{\'a}cs, G. 1981, Astrophysics and Space Science, 78, 175

\bibitem[{Leroy(2012)}]{leroy2012fastcalculation}
Leroy, B. 2012, Astronomy \& Astrophysics, 545, A50

\bibitem[{Leussu {et~al.}(2016)Leussu, Usoskin, Arlt, \&
  Mursula}]{leussu2016properties}
Leussu, R., Usoskin, I.~G., Arlt, R., \& Mursula, K. 2016, Astronomy \&
  Astrophysics, 592, A160

\bibitem[{Leussu {et~al.}(2017)Leussu, Usoskin, Pavai, Diercke, Arlt, Denker,
  \& Mursula}]{leussu2017wingsof}
Leussu, R., Usoskin, I.~G., Pavai, V.~S., {et~al.} 2017, Astronomy \&
  Astrophysics, 599, A131, publisher: EDP Sciences

\bibitem[{Li {et~al.}(2001)Li, Yun, \& Gu}]{li2001latitude}
Li, K.~J., Yun, H.~S., \& Gu, X.~M. 2001, The Astronomical Journal, 122, 2115,
  publisher: IOP Publishing

\bibitem[{Liu \& Nguyen(1998)}]{liu1998anaccurate}
Liu, Q. \& Nguyen, N. 1998, IEEE Microwave and Guided Wave Letters, 8, 18

\bibitem[{Lomb(1976)}]{lomb1976leastsquares}
Lomb, N.~R. 1976, Astrophysics and Space Science, 39, 447

\bibitem[{Mathias {et~al.}(2004)Mathias, Grond, Guardans, Seese, Canela,
  Diebner, \& Baiocchi}]{mathias2004algorithms}
Mathias, A., Grond, F., Guardans, R., {et~al.} 2004, Journal of Statistical
  Software, 11, 1

\bibitem[{McCracken {et~al.}(2013)McCracken, Beer, Steinhilber, \&
  Abreu}]{mccracken2013theheliosphere}
McCracken, K., Beer, J., Steinhilber, F., \& Abreu, J. 2013, Space Science
  Reviews, 176, 59

\bibitem[{Mortier {et~al.}(2015)Mortier, Faria, Correia, Santerne, \&
  Santos}]{mortier2015bglsa}
Mortier, A., Faria, J.~P., Correia, C.~M., Santerne, A., \& Santos, N.~C. 2015,
  Astronomy \& Astrophysics, 573, A101

\bibitem[{Munteanu {et~al.}(2016)Munteanu, Negrea, Echim, \&
  Mursula}]{munteanu2016effectof}
Munteanu, C., Negrea, C., Echim, M., \& Mursula, K. 2016, Annales Geophysicae,
  34, 437, publisher: Copernicus GmbH

\bibitem[{Ogurtsov {et~al.}(2002)Ogurtsov, Nagovitsyn, Kocharov, \&
  Jungner}]{ogurtsov2002longperiod}
Ogurtsov, M.~G., Nagovitsyn, Y.~A., Kocharov, G.~E., \& Jungner, H. 2002, 24

\bibitem[{Oppenheim(1999)}]{Oppenheim1999}
Oppenheim. 1999, Discrete-Time Signal Processing, 2nd edn. (Prentice Hall)

\bibitem[{Parzen(1962)}]{parzen1962stochastic}
Parzen, E. 1962, Stochastic Processes, ed. E.~L. Lehmann, Holden {Day} {Series}
  in {Probability} and {Statistics} (San Francisco: Holden-Day)

\bibitem[{Press \& Rybicki(1989)}]{PressRybicki1989}
Press, W.~H. \& Rybicki, G.~B. 1989, The Astrophysical Journal, 338, 277

\bibitem[{Prestes {et~al.}(2006)Prestes, Rigozo, Echer, \&
  Vieira}]{prestes2006spectral}
Prestes, A., Rigozo, N.~R., Echer, E., \& Vieira, L. E.~A. 2006, Journal of
  Atmospheric and Solar-Terrestrial Physics, 68, 182

\bibitem[{Scargle(1982)}]{scargle1982studies}
Scargle, J.~D. 1982, The Astrophysical Journal, 263, 835

\bibitem[{Seilmayer(2019)}]{seilmayer2019commonmethods}
Seilmayer, M. 2019, Common {Methods} of {Spectral} {Data} {Analysis}

\bibitem[{Seilmayer {et~al.}(2014)Seilmayer, Galindo, Gerbeth, Gundrum,
  Stefani, Gellert, R{\"u}diger, Schultz, \&
  Hollerbach}]{seilmayer2014experimental}
Seilmayer, M., Galindo, V., Gerbeth, G., {et~al.} 2014, Physical Review
  Letters, 113, 024505

\bibitem[{Seilmayer {et~al.}(2016)Seilmayer, Gundrum, \&
  Stefani}]{seilmayer2016}
Seilmayer, M., Gundrum, T., \& Stefani, F. 2016, Flow measurement and
  instrumentation, 48, 74

\bibitem[{Shannon(1949)}]{shannon1949communication}
Shannon, C.~E. 1949, Proceedings of the IRE, 37, 10

\bibitem[{Spoerer \& Maunder(1890)}]{spoerer1890profspoerers}
Spoerer, F. W.~G. \& Maunder, E.~W. 1890, Monthly Notices of the Royal
  Astronomical Society, 50, 251

\bibitem[{Spoerer(1889)}]{spoerer1889memoires}
Spoerer, G. 1889, Bulletin Astronomique, Serie I, 6, 60

\bibitem[{Sudars(2010)}]{sudars2010dataacquisition}
Sudars, K. 2010, Automatic Control and Computer Sciences, 44, 199

\bibitem[{Thompson \& Tree(1980)}]{thompson1980leakage}
Thompson, J.~K. \& Tree, D.~R. 1980, Journal of Sound and Vibration, 71, 531

\bibitem[{Townsend(2010)}]{townsend2010fastcalculation}
Townsend, R. H.~D. 2010, The Astrophysical Journal Supplement Series, 191, 247

\bibitem[{Usoskin(2017)}]{usoskin2017ahistory}
Usoskin, I.~G. 2017, Living Reviews in Solar Physics, 14, 3

\bibitem[{VanderPlas(2017)}]{vanderplas2017understanding}
VanderPlas, J.~T. 2017, The Astrophysical Journal, 236

\bibitem[{Weisstein(2019)}]{weisstein2019orthogonal}
Weisstein, E.~W. 2019, Orthogonal {Functions}

\bibitem[{Willis {et~al.}(2016)Willis, Wild, \&
  Warburton}]{willis2016reexamination}
Willis, D.~M., Wild, M.~N., \& Warburton, J.~S. 2016, Solar Physics, 291, 2519

\bibitem[{Zechmeister \& K{\"u}rster(2009)}]{zechmeister2009thegeneralised}
Zechmeister, M. \& K{\"u}rster, M. 2009, Astronomy \& Astrophysics, 496, 577

\end{thebibliography}

\appendix

\part*{Appendix}

\section{OMD as a non-consistent estimator}

\setcounter{equation}{0}
\numberwithin{equation}{section}

\label{sec:Orthogonal-mode-decomposition}In the following, only signals
$s(t)\in\mathbb{R}$ described in terms of a finite set of individual
frequency components $\omega_{k},0\leq k\leq M$ $(k\in\mathbb{N}^{+}),$
are considered. Furthermore, band limitation is assumed so that there
exists an upper maximum frequency $\omega_{k}<\omega_{\max}$.

The analysis is based on the trigonometric model definition from equation
(\ref{eq:QDT_model-1}) with the coefficients $a_{k},b_{k}\in\mathbb{R}$.
The model misfit $\epsilon(t)$ is defined by the difference between
the observed signal $s(t)$ and the assumed model function (\ref{eq:QDT_model-1})
as seen in equation (\ref{eq:model-misfit}). It follows Eq.~(\ref{eq:ModelWithError})
\begin{align*}
s(t) & =\sum_{k=0}^{M}\left(a_{k}\cos(\omega_{k}t)+b_{k}\sin(\omega_{k}t)+\epsilon_{k}(t)\right),
\end{align*}
the signal as sum of a trigonometric model with $M$ coefficients
and individual error values. The total model misfit $\epsilon(t)=\sum_{k}\epsilon_{k}(t)$
originates form measurement uncertainties with unknown distribution.

However, the sine and cosine functions are assumed as an orthonormal
basis, which is valid for the infinite integral, as seen in Section~\ref{subsec:Trigonometric-orthogonal-basis}.
To shorten the explanations the derivation refers only to the cosine
term and neglects the corresponding sine term, which can be always
achieved in similar manner.

To show the effects of OMD applied to discrete sampled functions,
e.\,g. taking time series from measurements, the next passage derives
the sampling series and its discrete model representation.

In general, a continuous signal can be described as a function defined
for every $-\infty<t<\infty$. But a realistic measurement or observation
$y(t)$ of such a process takes place in the range from $t=0$ to
arbitrary time $t=T$. Therefore, we associate to the measurement
a windowed signal $s_{\textrm{w}}(t)$ defined for every $-\infty<t<\infty$
as 
\begin{align}
s_{\mathrm{w}}(t) & =\left(y(t)+\epsilon(t)\right)\cdot w(t)\nonumber \\
\text{with }w(t) & =\begin{cases}
1 & 0\leq t<T\\
0 & \text{otherwise}
\end{cases}\label{eq:QDT_window}
\end{align}
 where the sampling error $\epsilon(t)$ is included. The window function
$w(t)$ ensures the finite observation time $0\leq t<T$ but leave
the infinite definition range untouched.

The sampling procedure, described in the next step, relies on the
Dirac distribution and its properties. The Dirac impulse is defined
as 
\begin{align}
\delta(t) & =\begin{cases}
\infty & t=0\\
0 & t\neq0
\end{cases}\text{ with }\int_{-\infty}^{\infty}\delta(t)\diff t=1.
\end{align}
The so-called sifting property
\begin{equation}
\int_{-\infty}^{\infty}\delta(t-\tau)\cdot\phi(t)\diff t=\phi(\tau),\label{eq:QDT_sifting}
\end{equation}
can be expressed for every function $\phi(t)$. Using this identity
the convolution function 
\begin{equation}
\Psi(t)=\sum_{n=-\infty}^{\infty}\delta(t-nT_{\mathrm{s}})
\end{equation}
helps to describe the sampling of the signal $s_{\mathrm{w}}(t)$
with sampling rate $T_{\mathrm{s}}$.

The continuous description of the sampling series 
\begin{equation}
A(t)=s_{\mathrm{w}}(t)\Psi(t)
\end{equation}
mathematically models the sampling which takes place while taking
a time series measurement of a physical process.

According to the theory of orthogonal function decomposition \textendash{}
briefly described and proven by \citet[Chap. 15]{cohen1995timefrequency}
\textendash{} the individual model coefficients of Eq. (\ref{eq:QDT_model-1})
\begin{align}
a_{k} & =\frac{2}{T}\int_{-\infty}^{\infty}y(t)\cos(\omega_{k}t)\diff t\label{eq:QDT_ak}\\
\text{and }b_{k} & =\frac{2}{T}\int_{-\infty}^{\infty}y(t)\sin(\omega_{k}t)\diff t\label{eq:QDT_bk}
\end{align}
can be recovered by integrating over the model function.

In order to estimate the coefficients $a_{k},b_{k}$ of the model
equation (\ref{eq:QDT_model-1}) the integration of (\ref{eq:QDT_ak})
and (\ref{eq:QDT_bk}) has to be carried out on the sampled series
$A(t)$ representing the acquired data. Assuming that the trigonometric
model $y(t)$ approximates the sampling series $A(t)$ in the limit
$\epsilon(t)\rightarrow0$: 
\begin{align}
a_{k} & =\frac{2}{T}\int_{-\infty}^{\infty}y(t)\cos(\omega_{k}t)\diff t\nonumber \\
 & \approx\frac{2}{T}\int_{-\infty}^{\infty}A(t)\cos(\omega_{k}t)\diff t\nonumber \\
 & =\frac{2}{T}\int_{-\infty}^{\infty}\left(y(t)+\epsilon(t)\right)w(t)\sum_{n=-\infty}^{\infty}\delta(t-nT_{\mathrm{s}})\cos(\omega_{k}t)\diff t,
\end{align}
and by factorizing and exchanging the sum with the integral
\begin{align}
a_{k} & =\frac{2}{T}\sum_{n=-\infty}^{\infty}\int_{-\infty}^{\infty}\underset{\text{sifiting properiy \eqref{eq:QDT_sifting}}}{\underbrace{\left(y(t)+\epsilon(t)\right)w(t)\cos(\omega_{k}t)\delta(t-nT_{\mathrm{s}})}}\diff t\nonumber \\
 & =\frac{2}{T}\sum_{n=-\infty}^{\infty}\left(y(nT_{\mathrm{s}})+\epsilon(nT_{\mathrm{s}})\right)w(nT_{\mathrm{s}})\cos(\omega_{k}nT_{\mathrm{s}}),
\end{align}
where the applied sifting property of the Dirac distribution achieves
the sampling at discrete time instances $nT_{\mathrm{s}}$. Due the
definition of the rectangular window function $w(t)$ the sum with
$nT_{\mathrm{s}}<0$ and $nT_{\mathrm{s}}>T$ is exactly zero, which
lead to a finite summation range. With the identity of the total number
of samples taken, $N=T/T_{\mathrm{s}}$, the coefficient $a_{k}$
reads as follows
\begin{align}
a_{k} & =\frac{2}{T}\sum_{n=0}^{N-1}\left(y(nT_{\mathrm{s}})+\epsilon(nT_{\mathrm{s}})\right)\cos(\omega_{k}nT_{\mathrm{s}})\nonumber \\
 & =\frac{2}{T}\sum_{n=0}^{N-1}\Biggl(\sum_{j=0}^{M}\bigl(a_{j}\cos(\omega_{j}nT_{s})\\
 & \phantom{=\frac{2}{T}\sum_{n=0}^{N-1}\Biggl(\sum_{j=0}^{M}\bigl(}+b_{j}\sin(\omega_{j}nT_{s})+\epsilon(nT_{\mathrm{s}})\bigr)\Biggr)\cos(\omega_{k}nT_{\mathrm{s}})\nonumber 
\end{align}

Figure~\ref{fig:Sinus-integration} sketches the scenario with $T_{\mathrm{s}}=f_{\mathrm{s}}^{-1}$
as the sampling period and $T>2\pi/\omega_{0}$. With respect to the
integrals (\ref{eq:QDT_ak}) and (\ref{eq:QDT_bk}) the dashed right
area causes errors in two ways, when QDT is carried out. First, the
energy of that amplitude is spread into the next neighboring integer
$k$s and second the truncation error (dashed area) causes a mismatch
of $a_{k}$ (and $b_{k}$) which only depends on $T$ but \emph{not}
on the amount of sampling points $N$ used.

In the following steps, the coefficients $k\neq j$ are neglected,
since they are projected in the error $\epsilon(t)$. The remaining
$k$-th set of parameters $k=j$ is sufficient to derive the concluding
points (i)-(iv) from Section~\ref{subsec:Trigonometric-orthogonal-basis}.

\subsection{Properties of the truncation error \textendash{} (i), (ii)}

Considering the Fourier decomposition of the sampling series $A(t)$
\begin{multline}
\int_{0}^{T}\sum_{n=0}^{N-1}s_{\mathrm{w}}(nT_{\mathrm{s}})\delta(t-nT_{\mathrm{s}})\cos(\omega_{k}nT_{\mathrm{s}})\diff t\\
=\int_{0}^{T}\delta\left(t-nT_{\mathrm{s}}\right)\diff t\sum_{n=0}^{N-1}a_{k}\cos^{2}(\omega_{k}nT_{\mathrm{s}})\\
+b_{k}\sin(\omega_{k}nT_{\mathrm{s}})\cos(\omega_{k}nT_{\mathrm{s}})+\epsilon_{k}\cos(\omega_{k}nT_{\mathrm{s}}),
\end{multline}
which can be rewritten with respect to the window function $w(t)$.
The latter enables the limitation of the integration and summation
boundaries. Next, by applying the sifting property (\ref{eq:QDT_sifting})
the continuous time series becomes independent from time so that
\begin{align}
\sum_{n=0}^{N-1}s_{\mathrm{w}}(nT_{\mathrm{s}})\cos(\omega_{k}nT_{\mathrm{s}}) & =\sum_{n=0}^{N-1}\left[a_{k}\cos^{2}(\omega_{k}nT_{\mathrm{s}})\right.\label{eq:diskrete_FS}\\
 & \phantom{=\sum_{n=0}^{N-1}}+b_{k}\sin(\omega_{k}nT_{\mathrm{s}})\cos(\omega_{k}nT_{\mathrm{s}})\nonumber \\
 & \phantom{=\sum_{n=0}^{N-1}}\left.+\epsilon_{k}\cos(\omega_{k}nT_{\mathrm{s}})\right]\nonumber 
\end{align}
describes the measured (sampled) data points at time instances $nT_{\mathrm{s}}$
of the signal. The expression above represents the well known sum
of the $k$-th cosine term of discrete Fourier series. The corresponding
sine term is defined in similar manner.

With respect to the trigonometric identities $\cos^{2}(x)=\frac{1}{2}\left(1+\cos\left(2x\right)\right)$
and with $\sin(x)\cos(x)=\frac{1}{2}\sin(2x)$ Equation~(\ref{eq:diskrete_FS})
becomes \begin{widetext}
\begin{align}
\sum_{n=0}^{N-1}s_{\mathrm{w}}(nT_{\mathrm{s}})\cos(\omega_{k}nT_{\mathrm{s}}) & =\sum_{n=0}^{N-1}\left(\frac{a_{k}}{2}\left(1+\cos(2\omega_{k}nT_{\mathrm{s}})\right)+\frac{b_{k}}{2}\sin(2\omega_{k}nT_{\mathrm{s}})+\epsilon_{k}\cos(\omega_{k}nT_{\mathrm{s}})\right)\\
\frac{2}{N}\sum_{n=0}^{N-1}s_{\mathrm{w}}(nT_{\mathrm{s}})\cos(\omega_{k}nT_{\mathrm{s}}) & =a_{k}\left(1+\frac{1}{N}\sum_{n=0}^{N-1}\left[\cos(2\omega_{k}nT_{\mathrm{s}})+\frac{b_{k}}{a_{k}}\sin(2\omega_{k}nT_{\mathrm{s}})\right]\right)+\frac{2}{N}\sum_{n=0}^{N-1}\epsilon_{k}\cos(\omega_{k}nT_{\mathrm{s}})\label{eq:diskrete_FS2}
\end{align}

\end{widetext}Assuming $b_{k}\approx0$, which corresponds to a pure
$\cos(x)$ signal the error can be divided into two components
\begin{align}
\epsilon_{\mathrm{T}} & =\frac{1}{N}\sum_{n=0}^{N-1}\cos(2\omega_{k}nT_{\mathrm{s}})\\
 & =\frac{1}{N}\Biggl(\underset{=0}{\underbrace{\sum_{n=0}^{N_{2\pi}}\cos(2\omega_{k}nT_{\mathrm{s}})}}+\sum_{n=N_{2\pi}+1}^{N-1}\cos(2\omega_{k}nT_{\mathrm{s}})\Biggl),\nonumber 
\end{align}
from which the first vanishes because it covers an integer number
of periods $\omega_{k}$. The right one consequently describes the
truncation error with respect to a $2\omega_{k}$ frequency as given
in the gray area in Figure \ref{fig:Sinus-integration}. A Taylor
series decomposition gives
\begin{align}
\epsilon_{\mathrm{T}} & =\frac{T_{\mathrm{s}}}{T}\sum_{n=N_{2\pi}+1}^{N-1}\Biggl(1-\frac{1}{2!}(2\omega_{k}nT_{\mathrm{s}})^{2}+\underset{\rightarrow0}{\underbrace{\frac{1}{4!}(2\omega_{k}nT_{\mathrm{s}})^{4}-\dots}}\biggl)\nonumber \\
 & \approx\frac{T_{\mathrm{s}}}{T}\left(\Delta N-2\left(\omega_{k}T_{\mathrm{s}}\right)^{2}\sum_{n=N_{2\pi}}^{N-1}n^{2}\right)
\end{align}
which is reduced further with $\sum_{n=1}^{N}n^{2}=\frac{N\left(N+1\right)\left(2N+1\right)}{6}$
and $T_{\mathrm{s}}=T/N$
\begin{align}
\epsilon_{\mathrm{T}} & \approx\frac{\Delta N}{N}-2\left(\omega_{k}T\right)^{2}\frac{\Delta N\left(\Delta N+1\right)\left(2\Delta N+1\right)}{6N^{3}}\nonumber \\
 & \approx\frac{\Delta N}{N}\left(1-2\left(\omega_{k}T\right)^{2}\left(\frac{2\Delta N^{2}}{6N^{2}}+\underset{\rightarrow0}{\underbrace{\frac{3}{6N^{2}}+\frac{1}{6N^{2}\Delta N}}}\right)\right)\nonumber \\
 & \approx\frac{\Delta N}{N}\left(1-\frac{2}{3}\left(\omega_{k}T\right)^{2}\left(\frac{\Delta N}{N}\right)^{2}\right)\qquad\text{with}\quad N=T/T_{\mathrm{s}}\nonumber \\
 & \approx\frac{\Delta\varphi}{T}\leq0.2\label{eq:QDT_res}
\end{align}
so that $\epsilon_{\mathrm{T}}$ becomes independent of the sampling
rate $T_{\mathrm{s}}$. The ``time phase'' $\Delta\varphi$ covers
the range marked as gray area in Figure~\ref{fig:Sinus-integration}.
The maximum value of $\epsilon_{\mathrm{T}}\leq0.2$ originates from
the $\cos(2\omega_{k}nT_{\mathrm{s}})$-term if only one period plus
truncation fits into the integration window. Moreover equation (\ref{eq:QDT_res})
implies, that a higher sampling frequency \textendash{} which gathers
more information from the process \textendash{} will not lead to a
more precise approximation of $a_{k}$ (and of course $b_{k}$). Therefore,
the QDT (or even the Fourier series decomposition) is \emph{not} a
consistent estimator for amplitude and phase, because it will not
converge
\begin{equation}
\lim_{N\rightarrow\infty}\sum_{n=0}^{N-1}y(nT_{\mathrm{s}})\cos(\omega_{k}nT_{\mathrm{s}})\neq a_{k}\label{eq:FS_convergence}
\end{equation}
towards the ``true'' $a_{k}$ for a given finite $T$. Instead of
that it converges in the limit of $T\rightarrow\infty$.

\subsection{Confidence intervals of model parameters \textendash{} (iii)}

The last term
\begin{align}
\frac{2}{N}\sum_{n=0}^{N-1}\epsilon_{k}\cos(\omega_{k}nT_{\mathrm{s}}) & =\frac{2}{N}\sum_{n=0}^{N-1}\mathcal{N}\left(0,\sigma\right)\cos(\omega_{k}nT_{\mathrm{s}})\nonumber \\
 & =\frac{2}{N}\Phi_{1-\alpha}\frac{\sigma}{\sqrt{N}}\sum_{n=0}^{N-1}\cos(\omega_{k}nT_{\mathrm{s}})\\
 & <\Phi_{1-\alpha}\frac{2\sigma}{\sqrt{N}}\nonumber \\
\epsilon_{\mathrm{FS}} & <\Phi_{1-\alpha}\frac{2\sigma}{\sqrt{N}}\label{eq:FS_random_error}
\end{align}
corresponds to the sampling error, which might be encountered in real
measurements. Given a normal distributed error function $\epsilon_{k}=\mathcal{N}\left(0,\sigma\right)$
with $\Phi_{1-\alpha}$ as the corresponding quantil, it turns out,
that this error suffice Equation~(\ref{eq:FS_random_error}) and
vanishes in the limit $N\rightarrow\infty$. This is valid because
a linear combination of normally distributed variables keeps normally
distributed, see Parzen \citet[Theorem 4A, p. 90]{parzen1962stochastic}.
And in addition, expression~(\ref{eq:FS_random_error}) gives the
upper limit of the parameter confidence interval.

In summary, the estimation of Fourier coefficients $a_{k}$ (and $b_{k}$)
\begin{equation}
\frac{2}{N}\sum_{n=0}^{N-1}s_{\mathrm{w}}(nT_{\mathrm{s}})\cos(\omega_{k}nT_{\mathrm{s}})\approx a_{k}\left(1+\frac{\Delta\varphi}{T}\right)\pm\Phi_{1-\alpha}\frac{2\sigma}{\sqrt{N}}
\end{equation}
is affected by the truncation error with respect to the full period
of $\omega_{k}$ and a random error from the measurement. The first
is independent from the sampling which proves that Fourier decomposition
is a non consistent estimator for amplitude and frequency. The last
term, the random error, converges to zero in the limit of large $N$,
as expected.

\pagebreak{}

\section{Table of Significant Periods}

\clearpage
\onecolumn

\begin{longtable}[t]{rrrrrl}
\caption{\label{tab:}\label{tab:Significant-peaks}Significant peaks of the sun spot spectrum}\\
\toprule
ID & $P_\mathrm{Yrs}/\mathrm{Yrs}$ & $\pm\Delta P/\mathrm{Yrs}$ & $\frac{f_\mathrm{Lat}}{\mathrm{10^{-2}\cdot deg}}$ & PSD & $\log_{10}(p)$\\
\midrule
1 & 559.540 & 216.060 & -3.170 & 0.001 & -23\\
2 & 358.855 & 266.266 & 2.946 & 0.001 & -34\\
3 & 162.651 & 505.535 & -1.225 & 0.004 & < -50\\
4 & 155.909 & 382.298 & 1.159 & 0.004 & < -50\\
5 & 124.893 & 142.836 & 5.426 & 0.001 & -29\\
\addlinespace
6 & 84.992 & 47.192 & 7.824 & 0.001 & -23\\
7 & 79.902 & 40.544 & 10.649 & 0.000 & -13\\
8 & 77.974 & 38.225 & -1.862 & 0.008 & < -50\\
9 & 75.367 & 35.247 & 1.453 & 0.005 & < -50\\
10 & 73.676 & 33.410 & -4.656 & 0.000 & -19\\
\addlinespace
11 & 73.497 & 33.219 & -12.487 & 0.000 & -13\\
12 & 68.258 & 27.982 & 3.916 & 0.001 & -48\\
13 & 65.974 & 25.892 & 13.117 & 0.000 & -15\\
14 & 51.097 & 14.731 & -1.593 & 0.006 & < -50\\
15 & 50.375 & 14.287 & 1.408 & 0.003 & < -50\\
\addlinespace
16 & 47.212 & 12.436 & -5.263 & 0.001 & -39\\
17 & 37.254 & 7.557 & 4.246 & 0.001 & -24\\
18 & 36.892 & 7.406 & -1.124 & 0.018 & < -50\\
19 & 36.798 & 7.366 & 1.525 & 0.011 & < -50\\
20 & 35.530 & 6.850 & -5.557 & 0.001 & -43\\
\addlinespace
21 & 34.615 & 6.490 & -11.245 & 0.000 & -18\\
22 & 32.852 & 5.826 & 9.391 & 0.000 & -15\\
23 & 29.882 & 4.795 & 1.386 & 0.008 & < -50\\
24 & 27.718 & 4.111 & 13.915 & 0.001 & -26\\
25 & 27.203 & 3.956 & -4.189 & 0.004 & < -50\\
\addlinespace
26 & 27.126 & 3.934 & 6.638 & 0.001 & -26\\
27 & 26.750 & 3.823 & 3.881 & 0.001 & -43\\
28 & 26.408 & 3.724 & -1.324 & 0.055 & < -50\\
29 & 26.283 & 3.688 & 1.207 & 0.044 & < -50\\
30 & 23.924 & 3.046 & 18.856 & 0.001 & -27\\
\addlinespace
31 & 23.816 & 3.018 & -18.655 & 0.000 & -11\\
32 & 21.684 & 2.495 & -1.082 & 0.600 & < -50\\
33 & 21.637 & 2.484 & 1.152 & 0.607 & < -50\\
34 & 21.284 & 2.403 & -6.092 & 0.004 & < -50\\
35 & 16.976 & 1.522 & 1.965 & 0.003 & < -50\\
\addlinespace
36 & 16.936 & 1.514 & -2.713 & 0.003 & < -50\\
37 & 15.424 & 1.254 & 1.160 & 0.018 & < -50\\
38 & 15.363 & 1.244 & -1.365 & 0.017 & < -50\\
39 & 13.123 & 0.906 & -3.694 & 0.001 & < -50\\
40 & 12.836 & 0.867 & 1.075 & 0.016 & < -50\\
\addlinespace
41 & 12.819 & 0.865 & -1.303 & 0.017 & < -50\\
42 & 10.668 & 0.598 & 2.450 & 0.002 & < -50\\
43 & 10.275 & 0.555 & -3.534 & 0.001 & < -50\\
44 & 10.106 & 0.537 & 0.982 & 0.003 & < -50\\
45 & 9.495 & 0.473 & 1.616 & 0.002 & < -50\\
\addlinespace
46 & 8.994 & 0.425 & -3.782 & 0.002 & < -50\\
47 & 8.980 & 0.423 & -1.204 & 0.008 & < -50\\
48 & 7.711 & 0.312 & -0.948 & 0.012 & < -50\\
49 & 7.704 & 0.311 & 1.414 & 0.010 & < -50\\
50 & 7.207 & 0.272 & 1.210 & 0.026 & < -50\\
\addlinespace
51 & 7.195 & 0.272 & -1.164 & 0.026 & < -50\\
52 & 7.181 & 0.271 & 2.910 & 0.021 & < -50\\
53 & 6.375 & 0.213 & -2.583 & 0.004 & < -50\\
54 & 5.919 & 0.184 & 2.345 & 0.003 & < -50\\
55 & 5.891 & 0.182 & -2.916 & 0.002 & < -50\\
\addlinespace
56 & 5.638 & 0.167 & 2.907 & 0.002 & < -50\\
57 & 4.558 & 0.109 & -2.872 & 0.003 & < -50\\
58 & 4.310 & 0.097 & 3.558 & 0.002 & < -50\\
59 & 4.247 & 0.095 & -2.963 & 0.003 & < -50\\
60 & 4.125 & 0.089 & -3.366 & 0.002 & < -50\\
\addlinespace
61 & 4.065 & 0.087 & 3.815 & 0.002 & < -50\\
62 & 3.944 & 0.081 & -3.364 & 0.002 & < -50\\
\bottomrule
\end{longtable}
\clearpage
\twocolumn
\end{document}